\documentclass[twocolumn,twocolappendix]{aastex6}
\usepackage{mathrsfs}

\begin{document}

\title{Revising the evolutionary stage of HD\,163899: the effects of convective overshooting and rotation}

\author{Jakub Ostrowski\altaffilmark{1,2}, Jadwiga Daszy\'nska-Daszkiewicz\altaffilmark{1} and Henryk Cugier\altaffilmark{1}}
\affil{Instytut Astronomiczny, Uniwersytet Wroc{\l}awski \\
ul. Kopernika 11 \\
51-622 Wroc{\l}aw, Poland}

\altaffiltext{1}{Uniwersytet Wroc{\l}awski}
\altaffiltext{2}{ostrowski@astro.uni.wroc.pl}

\begin{abstract}

We revise the evolutionary status of the B-type supergiant HD\,163899 based on the new determinations of the mass-luminosity ratio, effective temperature and rotational velocity as well as on the interpretation of the oscillation spectrum of the star. The observed value of the nitrogen-to-carbon abundance fixes the value of the rotation rate of the star. Now, more massive models are strongly preferred than those previously considered and it is very likely that the star is still in the main sequence stage. The rotationally induced mixing manifests as the nitrogen overabundance in the atmosphere, which agrees with our analysis of the HARPS spectra. Thus, HD\,163899 belongs probably to a group of evolved nitrogen-rich main-sequence stars. 

\end{abstract}


\section{Introduction} \label{introduction}

HD\,163899, a B2 Ib/II supergiant, was proposed by \citet{2006ApJ...650.1111S} as a prototype candidate of a new class of pulsating stars termed Slowly Pulsating B-type supergiants (SPBsg). The star was analysed by many groups  \citep{2006ApJ...650.1111S, 2008CoAst.157..311G,2013MNRAS.432.3153D,2015MNRAS.447.2378O} but all of these studies share the same drawback, which is the lack of proper determination of the basic stellar parameters. The spectral classification of HD\,163899 was done on the basis of photometric indices by \citet{1977A&AS...27..215K} and \citet{1996ApJS..104..101S} and the very crudely estimated basic parameters. Therefore, the main problem was establishing the evolutionary stage of the star. \citet{2006ApJ...650.1111S} and \citet{2009MNRAS.396.1833G} claimed it was a supergiant during the first crossing towards the red giant branch (RGB). \citet{2008CoAst.157..311G} suggested that HD\,163899 could be a main sequence star but lower masses derived from the old estimations of stellar parameters did not support this statement. In such case the enormously large value of overshooting from the convective core was required. In our previous paper \citep{2015MNRAS.447.2378O} we studied the possibility that HD\,163899 is a supergiant on the blue loop in the core helium burning phase. We found that for certain values of the convective overshooting parameter and metallicity,  blue loops can reach early B spectral types for stars with masses in the range of ~14 - 20 $M_\odot$, but such models could not reproduce the observed frequency range and the number of detected modes. Therefore, we concluded that the star would rather be in the phase of shell hydrogen burning. In this paper, we determine for the first time the basic parameters from high-resolution spectra. We also derive the value of the projected rotational velocity, $V_{\rm rot}\sin i$, which was not available in the previous studies. Moreover, we discuss the measured abundances of CNO elements and use them to apply constraints on parameters of the models, especially the rotation rates. Stellar rotation is an important source of element mixing and together with other mixing mechanisms (convection, semiconvection, convective overshooting) can greatly affect the evolution and internal structure of the star as well as observed surface abundances. Here, we focus on these issues and study their effects on stellar models using MESA code.

The values of convective overshooting studied here, $f \leq 0.03$, are fairly typical among the values considered in modern stellar modelling or calibrations with observational data (e.g. \citealt{2015A&A...575A.117S}). The grids of \citet{2012A&A...537A.146E} and \citet{2011A&A...530A.115B} use a step overshooting instead of exponential formula adopted here (eq.\,\ref{eq_overshooting}, \citealt{2000A&A...360..952H}). \citet{2012A&A...537A.146E} uses $\alpha_\mathrm{ov} = 0.1$, which roughly corresponds to $f = 0.01$ ($\alpha_\mathrm{ov} \approx 10f$, \citealt{2015EAS....71..317M}). The value of $f=0.03$, which gives good results in our calculations, is rather high but comparable to the value adopted by \citet{2011A&A...530A.115B}: $\alpha_\mathrm{ov} = 0.335$. They also adopted much more efficient semiconvective mixing with $\alpha_\mathrm{SC} = 1.0$, using the same prescription of \citet{1983A&A...126..207L} as we do. On the other hand, \citet{2012A&A...537A.146E} utilize Schwarzschild criterion for convective instability and hence do not take semiconvection into account. Recently, \citet{2015A&A...580A..27M} tried to build a seismic model to reproduce 19 frequencies of the SPB star KIC 10526294 using \texttt{MESA} models. They obtained the best fit for the models with overshooting of $f=0.017 - 0.018$, however the mass of the star was in the range of $[3.15, 3.25]~M_\odot$, which is significantly lower than the stars we study. More interesting insight about the value of overshooting can be found in the seismic modelling of two hybrid $\beta$ Cephei/SPB pulsators: 12 Lacertae \citep{2013MNRAS.431.3396D} and $\gamma$ Pegasi \citep{2013MNRAS.432..822W}. In both cases the models that fit the observed frequencies can be found within a quite wide range of overshooting parameter ($\alpha_\mathrm{ov} \sim 0.2 - 0.45$) because obviously other factors such as metallicity or opacities affect seismic models. In this paper, we choose the value of the core overshooting parameter of $f = 0.02$ and $0.03$, which reasonably lies within the accepted range.

Stellar pulsations are potentially powerful probes of stellar interior. They can yield constraints on parameters of stellar models and theory. For self-excited pulsation certain conditions have to be fulfilled and hence the changes in the internal structure at some stages of evolution can prevent a propagation of pulsational modes. For a long time it was thought that g-mode pulsations can not be excited in massive B-type stars after the terminal age main sequence (TAMS; \citealt{1999AcA....49..119P,2007CoAst.150..207P}). The reason was strong radiative damping expected in a dense radiative helium core where the Brunt-V\"ais\"al\"a frequency reaches huge values. The situation changed when \citet{2006ApJ...650.1111S} identified 48 frequency peaks below 3 d$^{-1}$ and with the maximum amplitude of a few millimagnitudes from the data obtained by the {\it MOST} satellite for HD\,163899 (B2 Ib/II supergiant). The authors attributed these frequencies to the g- and p-mode pulsations and explained their existence by a partial reflection of some eigenmodes at an intermediate convective zone (ICZ) related to the hydrogen-burning shell. The instability analysis performed by other groups \citep{2008CoAst.157..311G,2013MNRAS.432.3153D,2015MNRAS.447.2378O} confirmed that both p- and g-modes can be excited in the post main sequence massive stellar models by the $\kappa$-mechanism acting in the metal opacity bump. Thus, HD\,163899 has become a prototype of SPBsg stars. Other candidates have been also proposed (e.g. \citealt{2007A&A...463.1093L}). In this paper, for the first time, pulsational properties of HD\,163899 are confronted with rotating models calculated for a well established set of stellar parameters and rotational velocity derived from high-resolution spectroscopy.

The structure of this paper is as follows. In Section \ref{hd163899}, we present the new determinations of basic stellar parameters for HD\,163899. The MESA evolutionary models are presented in Section \ref{models}, where we discuss the influence of convective overshooting and stellar rotation on the internal structure and surface composition of the star. In Section \ref{frequencies}, we present pulsational calculations and interpret the observed and theoretical frequency distributions obtained for HD\,163899. The last section contains conclusions.

\section{Stellar parameters of HD\,163899} \label{hd163899}

\begin{figure*}
\begin{center}
 \includegraphics[clip,width=42mm,angle=0]{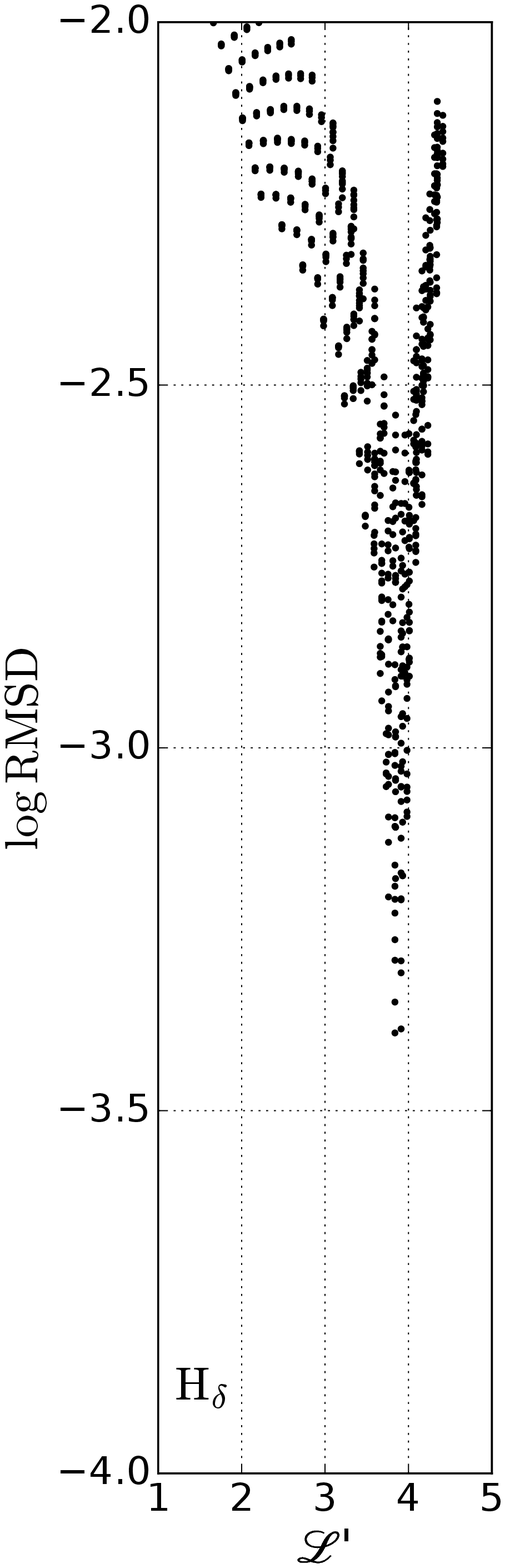}
 \includegraphics[clip,width=42mm,angle=0]{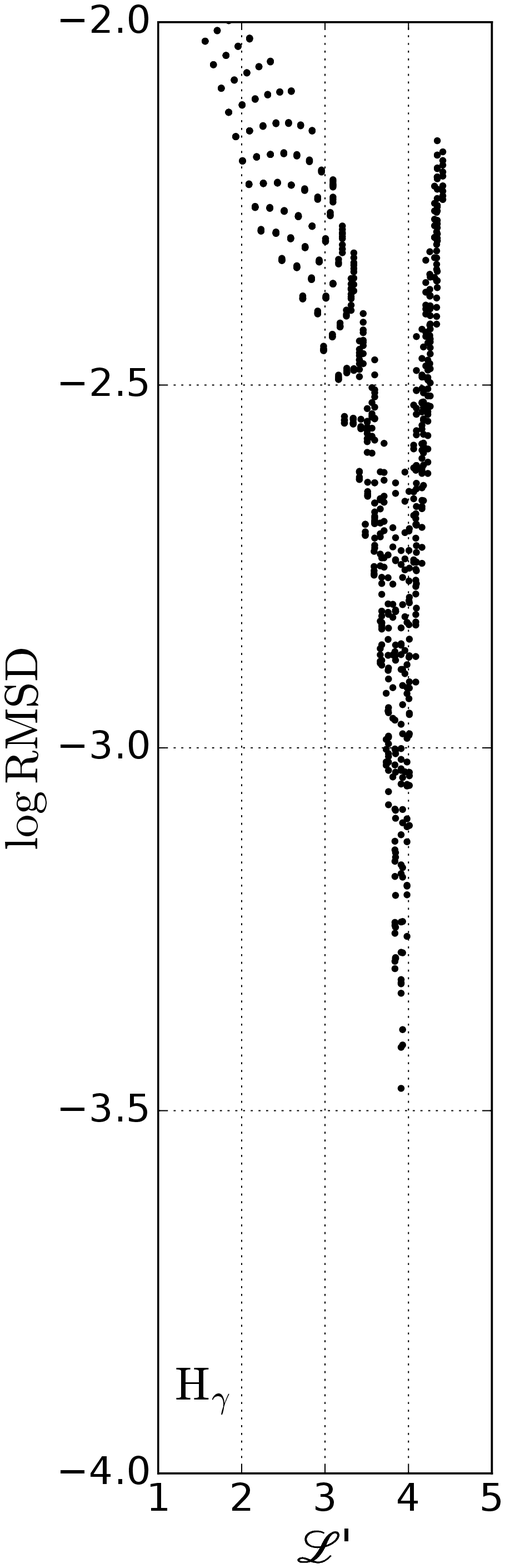}
 \includegraphics[clip,width=42mm,angle=0]{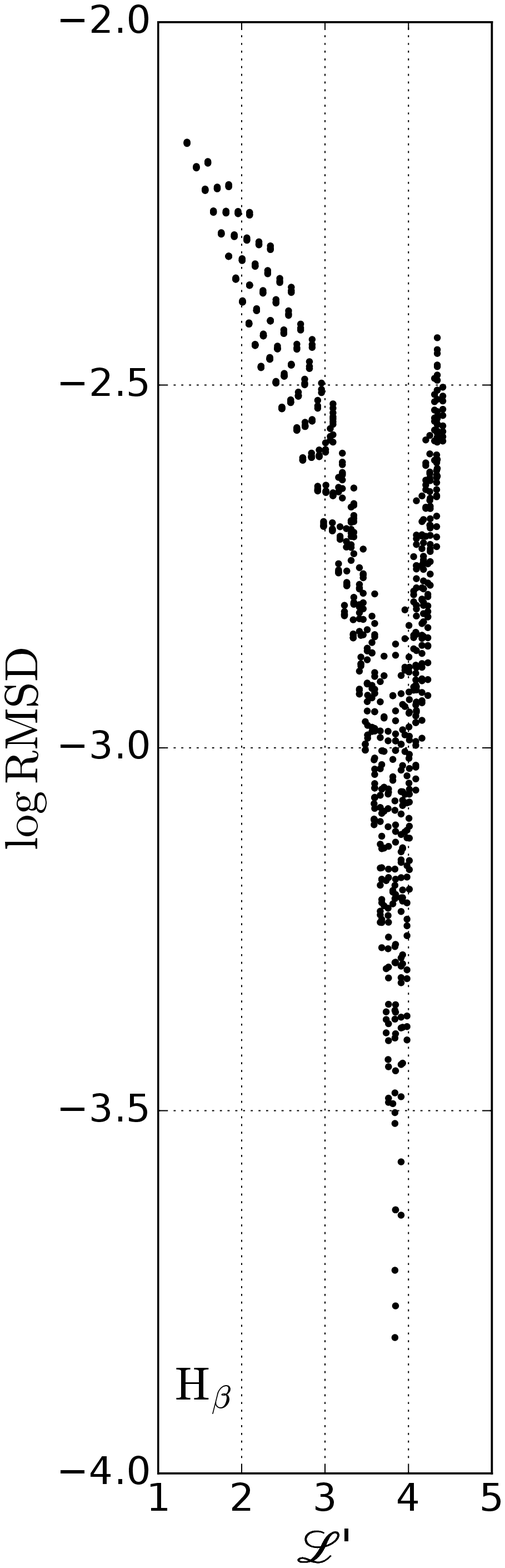}
 \includegraphics[clip,width=42mm,angle=0]{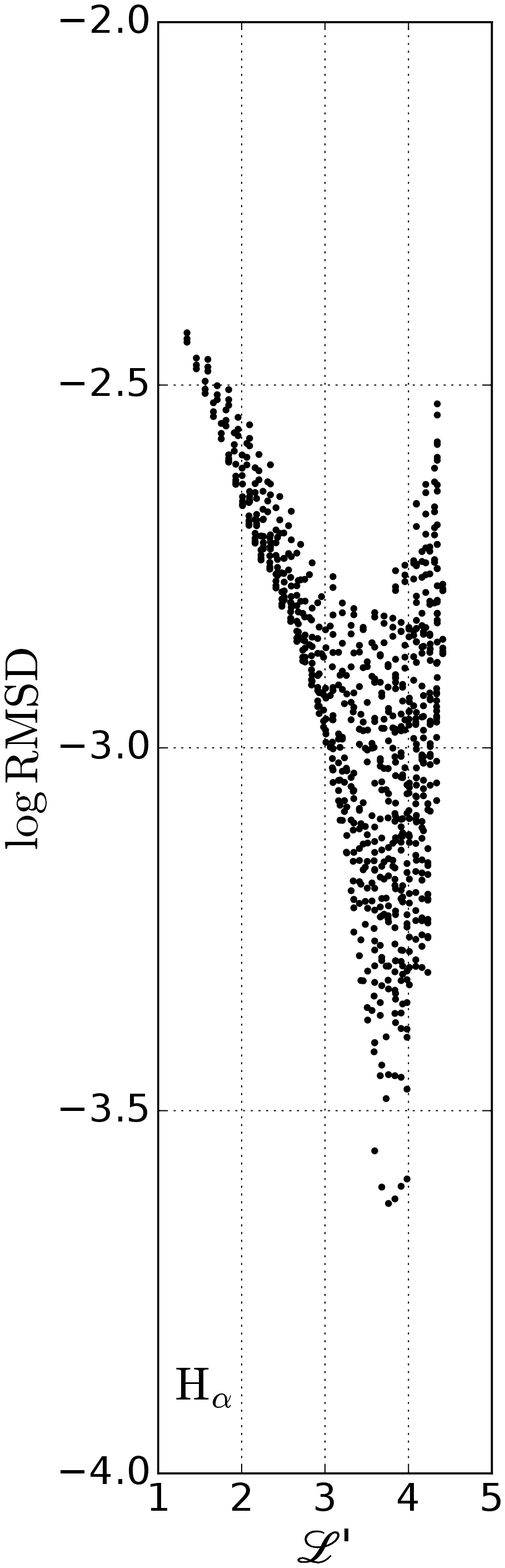}
 \caption{Logarithmic values of the root mean square deviation (RMSD) as a function of the parameter $\mathscr{L}^{'}$ for the four hydrogen lines. The best fit solutions correspond to the minimum value of RMSD obtained for the solar chemical abundance (GS98) and the microturbulent velocity $\xi_t=2$ $\mathrm{km~s^{-1}}$.}
\label{fig1}
\end{center}
\end{figure*}

\begin{figure*}
\begin{center}
\includegraphics[width=88mm,height=62.4mm,angle=0]{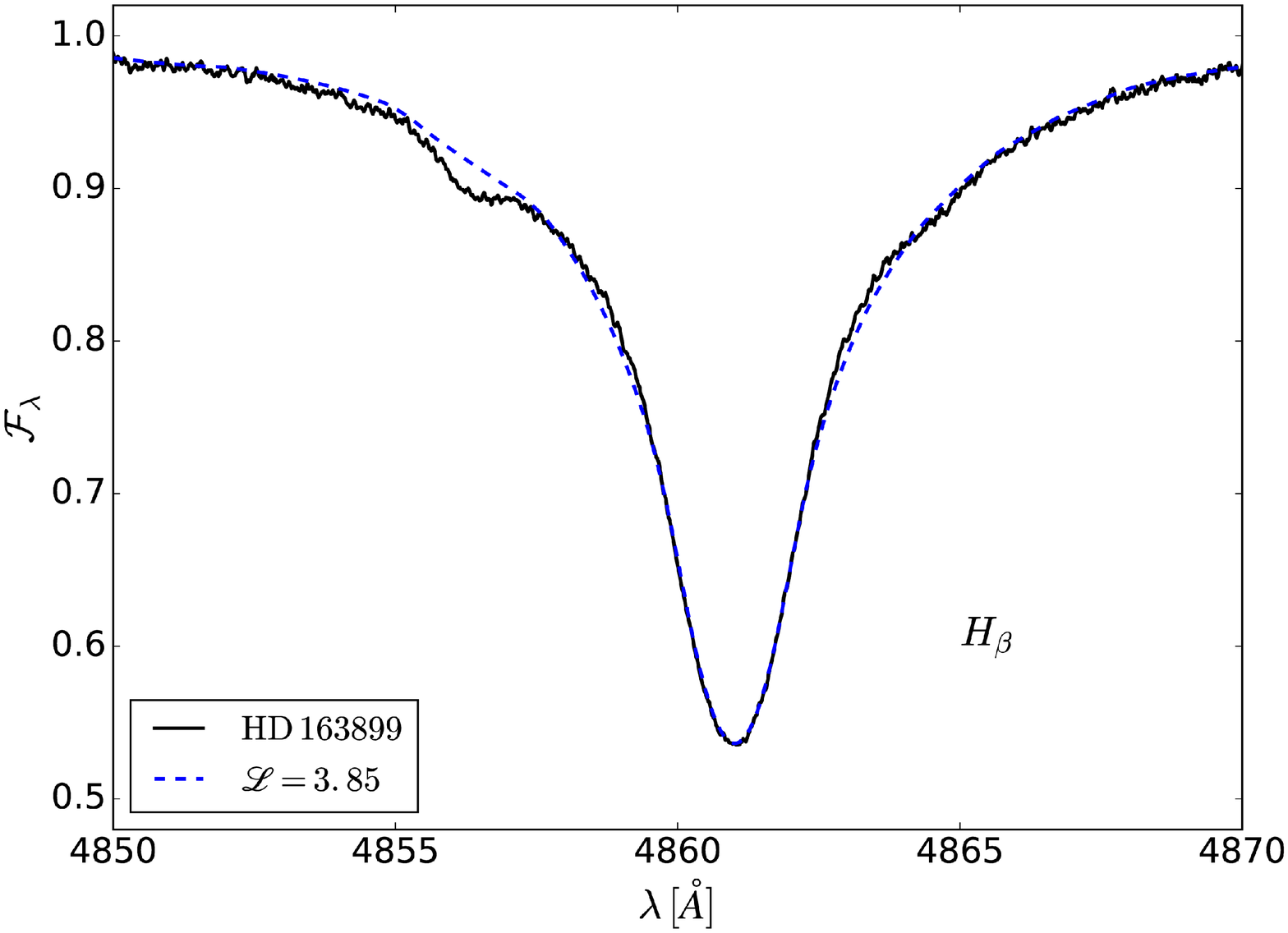}
\includegraphics[width=88mm,height=62.4mm,angle=0]{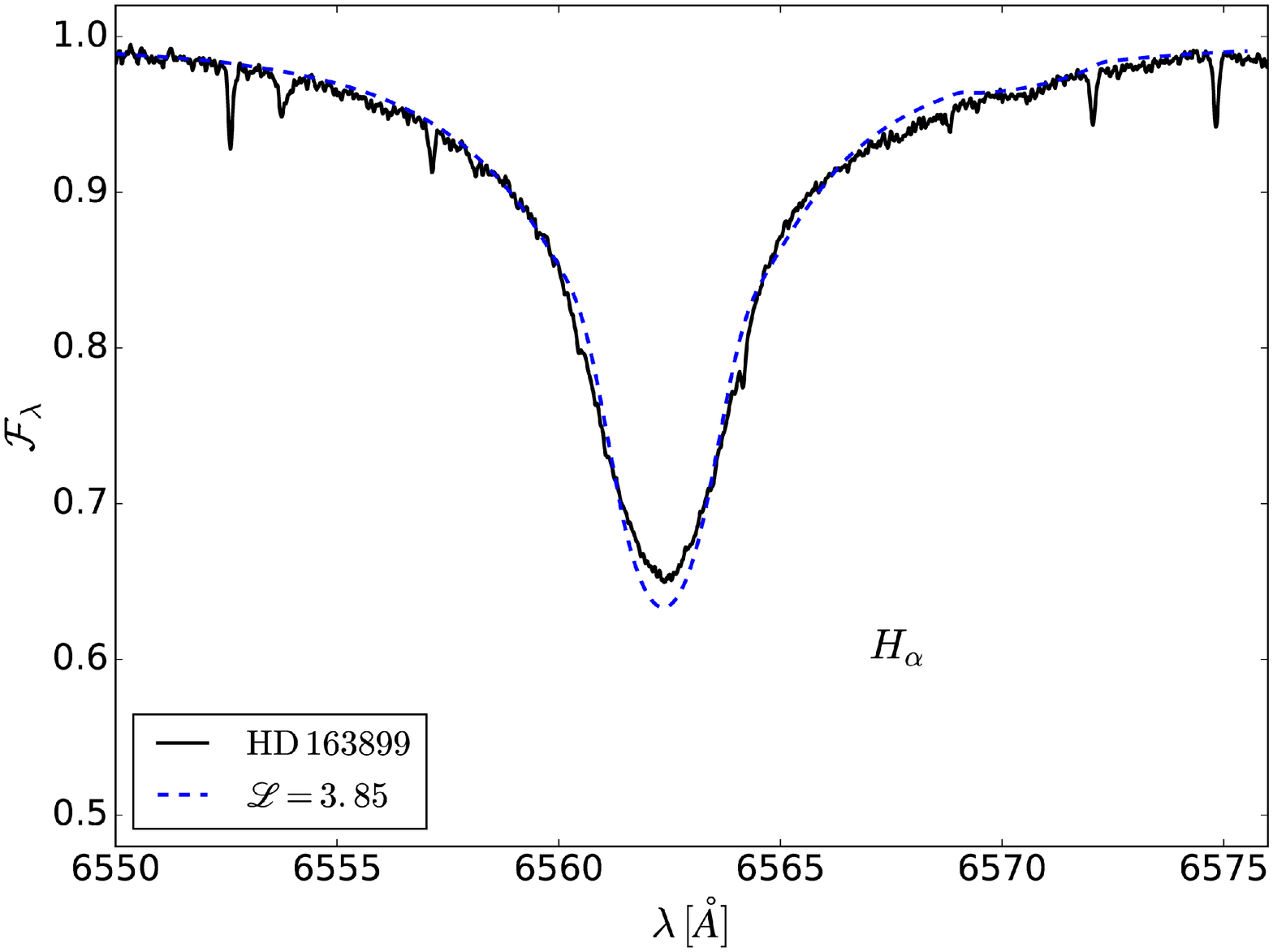}
\includegraphics[width=88mm,height=62.4mm,angle=0]{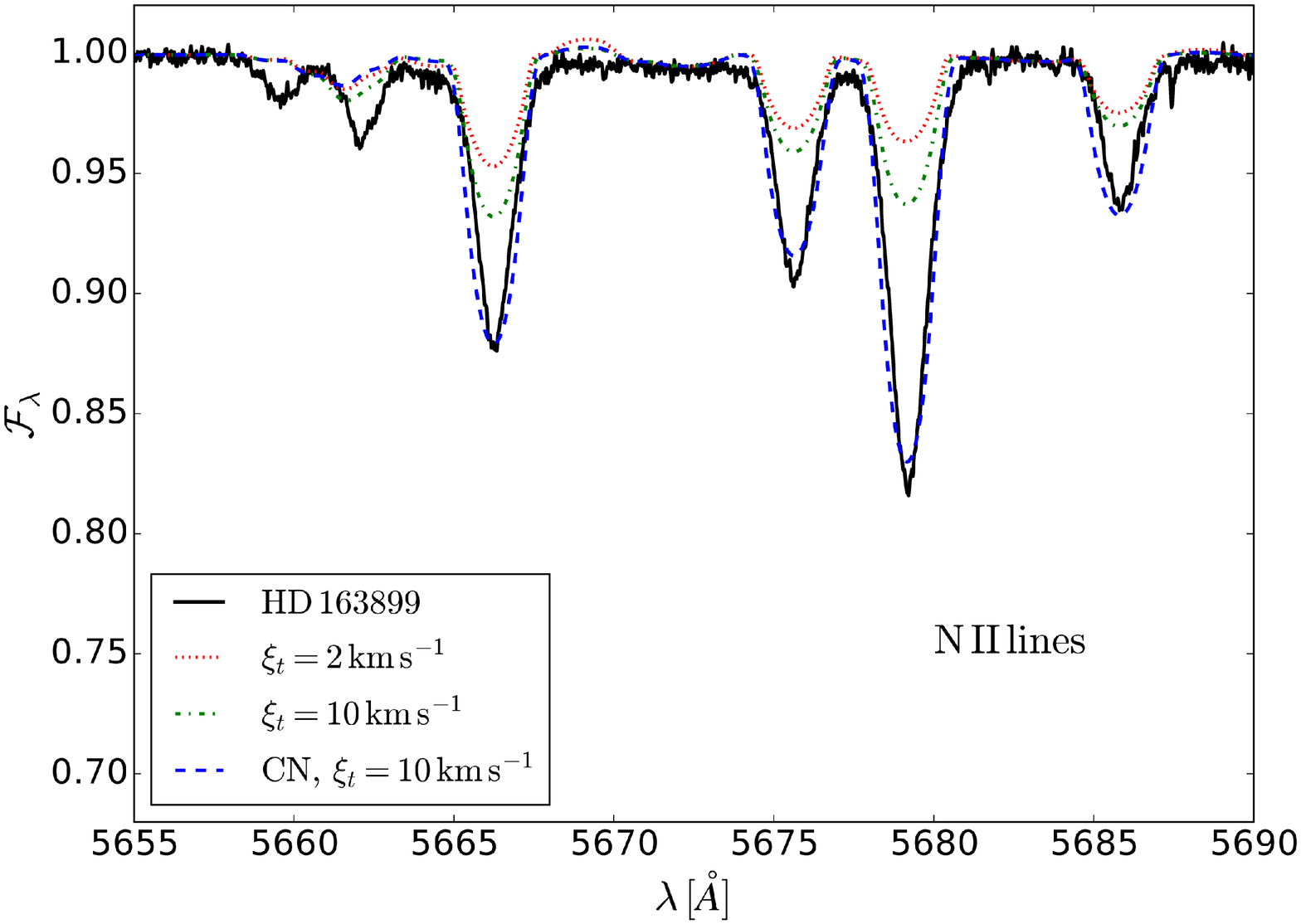}
\includegraphics[width=88mm,height=62.4mm,angle=0]{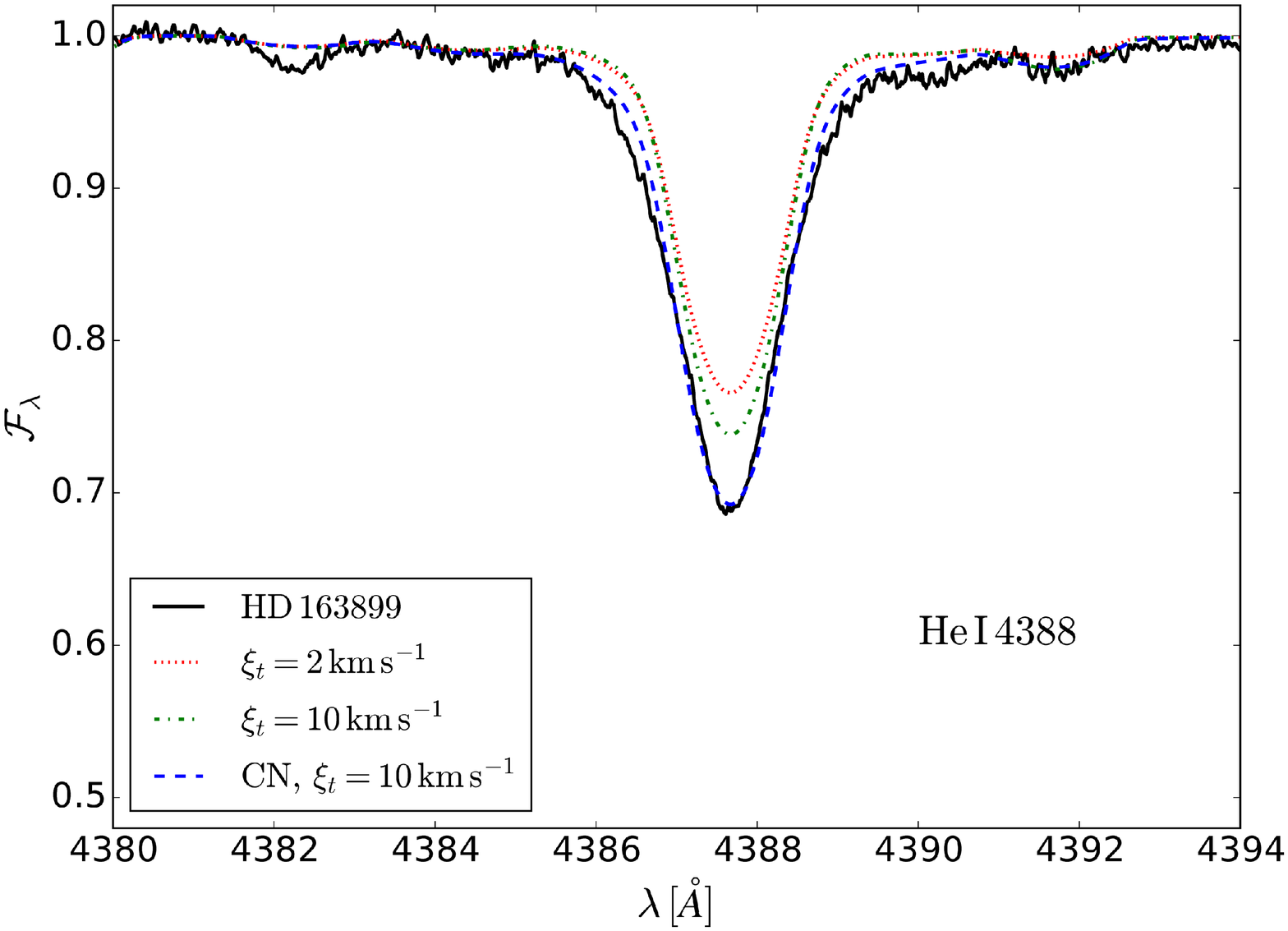}
  \caption{The observed mean spectrum of HD163899 (black lines) in comparison with synthetic spectra near the lines  H$_\beta$, H$_\alpha$, NII and  HeI\,4388 computed for $\mathscr{L}^{'} = 3.85$. In the bottom panels, the CN models are shown to present the effects of the microturbulent velocity, $\xi_t$, and an enrichment of nitrogen and helium. The strong nitrogen lines at the wavelength region $5660 - 5690$ {\AA} indicate an overabundance of this element in HD\,163899.}
\label{fig2}
\end{center}
\end{figure*}

\begin{table*}
\caption{The logbook of spectroscopic observations of HD\,163899 with the HARPS instrument.}
\label{table1}      
\centering                          
\begin{tabular}{c c c c c c c }        
\hline\hline                 
 No. & Spectrum             &  MJD          & Exp. time     & S/N  & $\lambda$  \\
     &                      & [d]           &  [s]          &      &  [{\AA}]   \\
\hline                        
 1 & HD163899-20110625-2320 & 2455738.48708 & 600     & 141 & 3785 - 6910   \\
 2 & HD163899-20120625-0053 & 2456103.54619 & 600     & 136 & 3785 - 6910   \\
 3 & HD163899-20120625-0103 & 2456103.55345 & 1600    & 212 & 3785 - 6910   \\
\hline
\end{tabular}
\end{table*}

\begin{table*}
\caption{Non-LTE atmosphere models used in the analysis. The BG and CN models were calculated by \citet{2007ApJS..169...83L} 
while the T1 - T4 models were calculated with the updated atom model of carbon (see text) in this paper.}             
\label{table2}      
\centering                          
\begin{tabular}{c c c c c c c c c c }        
\hline\hline                 
 No. & Model             &  $T_{\rm{eff}}$  & $\log g$  & $\xi_{\rm{t}}$                   &  &  & Abundances &     &  \\
     &                      & [K]           & [c.g.s] & [$\rm{km}$ $\rm{s^{-1}}$] & He/H & C/H & N/H & O/H &  \\
\hline                        
 1 & BG & 15000 - 30000 & 2.50 - 4.75 & 2,10 & 0.10      & $3.31$ $10^{-4}$ & $8.32$ $10^{-5}$     & $6.76$ $10^{-4}$ & GS98 \\
 2 & CN & 15000 - 30000 & 2.50 - 3.00 & 10   & 0.20      & $1.65$ $10^{-4}$ & $4.16$ $10^{-4}$     & $6.76$ $10^{-4}$ & GS98 \\
 3 & T1 & 15000 - 24000 & 3.00 - 3.75 & 2-20 & 0.085     & $3.28$ $10^{-4}$ & $8.24$ $10^{-5}$     & $6.76$ $10^{-4}$ & GS98 \\
 4 & T2 & 20000 - 25000 & 3.50        & 2-15 & 0.085     & $2.69$ $10^{-4}$ & $6.76$ $10^{-5}$     & $4.90$ $10^{-4}$ & AGSS09 \\
 5 & T3 & 23000         & 3.00        & 15   & 0.085     & $3.28$ $10^{-4}$ & (1.65-2.23)$10^{-4}$ & $6.70$ $10^{-4}$ & GS98 \\
 6 & T4 & 23000         & 3.00        & 15   & 0.15,0.20 & $3.28$ $10^{-4}$ & $1.83$ $10^{-4}$     & $6.70$ $10^{-4}$ & GS98 \\
\hline
\end{tabular}
\end{table*}

HD\,163899 has aroused more interest when \citet{2006ApJ...650.1111S} detected 48 frequencies in the {\it MOST} light curve. These light variations were explained by the existence of g- and p-mode pulsations. However, an obstacle in the interpretation of the HD\,163899 oscillations was the problem with determination of the evolutionary stage of the star because of the uncertain values of the basic parameters such as effective temperature or surface gravity.

Recently, HD\,163899 was a target object in a SPACEINN spectroscopic project, which supported the {\it MOST} satellite mission. The spectral type of B2 II, effective temperature, $T_{\rm{eff}} = 23961$ K, surface gravity, log g = 3.1, projected rotational velocity, $V \sin i = 48-49$ km/s and metallicity, [Fe/H] = 0.05 dex, were assigned for HD\,163899 in the SPACEINN archive data (http://www.brera.inaf.it/speceinn/data/). This is in good agreement with earlier classifications of this star as B2 I-II object based on photometric measurements \citep{1977A&AS...27..215K,1996ApJS..104..101S}. According to \citet{1977A&AS...27..215K}, $\beta = 2.577 \pm 0.006$ and $E(B-V) \approx 0.45$ mag. The reddening free parameter, $Q = (U-B) - E(U-B)/E(B-V) (B-V)$, is about $Q \approx -0.81$ mag and the usage of the $\beta$ vs. $Q$ diagram indicates the the star is close to the upper border of the region where $\beta$ Cephei stars are located \citep{1993SSRv...62...95S}.

Here, we analyse three high-resolution, high signal-to-noise spectra at the wavelength region of $\lambda = 3785 - 6910$ {\AA}. The logbook is given in Table\,\ref{table1}. The spectra are from the HARPS instrument \citep{2003Msngr.114...20M} on the 3.6-metre telescope at ESO's La Silla Observatory. The barycenter corrections for the wavelength scale have been already introduced. The interstellar Ca II K and H resonance lines reveal three absorption components at the radial velocities of -6.20, -12.35 and -21.51 $\mathrm{km~s^{-1}}$. The main component occurs at -6.20 $\mathrm{km~s^{-1}}$ with the central depths equal to zero. Telluric lines ($\mathrm{H_2 O}$ and $\mathrm{O_2}$) dominate at the long wavelength side of the observed spectra. They show exactly the same wavelengths in the three analyzed spectra, which indicates that the barycenter corrections were calculated with good precision. The same is true for the interstellar lines. Finally, the two analyzed spectra (Nos. 2 and 3 in Table\,\ref{table1}) are very close in time of observations and are similar to each other, while the third one (No. 1 obtained 365 days earlier) shows additional absorption features on the blue side of the line cores of $\mathrm{H}_\alpha$ - $\mathrm{H}_\gamma$. The depths of these features are less than 2.5 per cent, the radial velocities are about -45 $\mathrm{km~s^{-1}}$ and they can be related to variable stellar wind, pulsations or binarity. Given the limited spectroscopic data at our disposal, we cannot exclude the possibility that HD\,163899 has a pulsating lower mass companion.

In the next step, the observed spectra were compared with non-LTE atmosphere models listed in Table\,\ref{table2}. First, the synthetic spectra calculated by \citet{2007ApJS..169...83L} (LH) were broadened for the projected  rotational velocities of $V_{\mathrm{rot}} \sin i = 20, 40, 50, 60, 70 $ and $80$ $\rm{km}$ $\rm{s^{-1}}$. The spectra were also convolved with an instrumental profile and then interpolated to a denser grid of $T_{\mathrm{eff}}$, $\log g$ and $V_{\mathrm{rot}} \sin i$. Rather than separately determining the effective temperature, $T_{\mathrm{eff}}$, and gravity, $\log g$, as in a traditional spectroscopic analysis, we made use of the parameter $\mathscr{L}^{'} = 4 \log T_\mathrm{eff} - \log g - 10.609$, which can be uniquely assigned to each spectrum. The values of $\mathscr{L}^{'}$ were measured from the four hydrogen lines $\mathrm{H}_\delta, \mathrm{H}_\gamma, \mathrm{H}_\beta$ and $\mathrm{H}_\alpha$. The best fits correspond to the minimum value of the root mean square deviation (RMSD) and its logarithmic values as a function of $\mathscr{L}^{'}$ are plotted in Fig.\,\ref{fig1}. As one can see, there is good agreement for all studied hydrogen lines. However, $\mathrm{H}_\beta$ is most suitable for our purposes, because of the smallest blending effect by metallic lines. The quality of the solution can be judged from Fig.\,\ref{fig2} where we show an example of the fit of the theoretical spectra for $\mathscr{L}^{'} = 3.85$ to the two hydrogen lines, $\mathrm{H}_\beta$ and $\mathrm{H}_\alpha$ (the top panels), the helium line He I 4388 (the bottom-right panel) and the region with strong nitrogen lines (the left-right panel). To reproduce the He I and N II line profiles not only did we have to increase the microturbulent velocity from 2 to 10 $\mathrm{km~s^{-1}}$, but also we had to use atmosphere models enriched in nitrogen and helium. These are the CN models, in the notation of \citet{2007ApJS..169...83L}, which have the helium abundance increased to He/H = 0.2 by number, the nitrogen abundance increased by a factor of 5, and the carbon abundance halved. These spectral features are signatures of a surplus of helium and, in particular, of an overabundance of nitrogen in HD\,163899.

\begin{figure}
\begin{center}
\includegraphics[clip,width=87mm,angle=0]{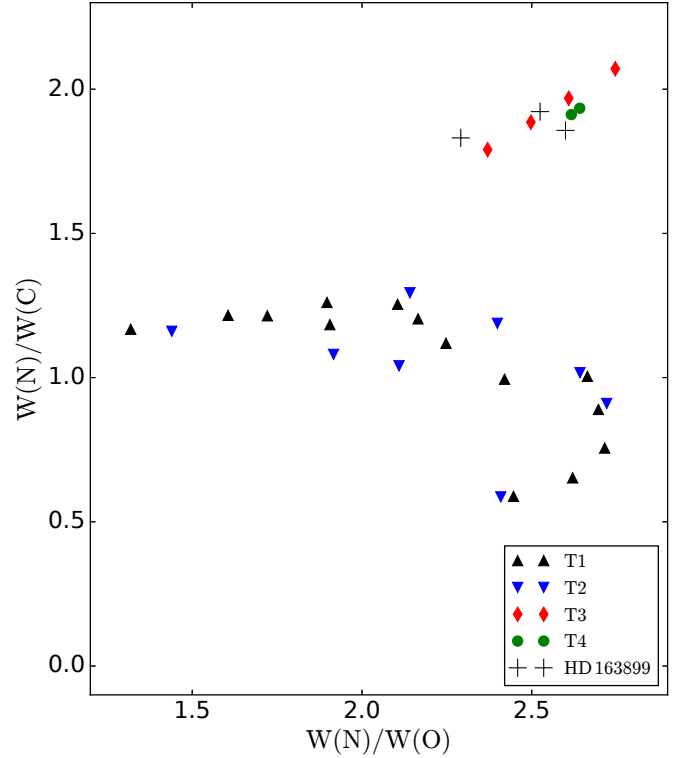}
  \caption{Ratios of the observed equivalent widths of W(N)/W(C) vs. ratios of W(N)/W(O) for selected lines of C II, N II and O II (see text, Sec.\,\ref{hd163899}) in comparison with the T atmosphere model (Table\,\ref{table2}) predictions.}
\label{fig3}
\end{center}
\end{figure}

The overabundance of nitrogen for HD\,163899 is also seen in the diagram W(N)/W(C) vs. W(N)/W(O), showed in Fig.\,\ref{fig3}, where W(C), W(N) and W(O) mean the equivalent widths of carbon lines at 5132.947, 5133.282, 5139.174, 5143.495, 5145.165 and 5151.085 {\AA}, nitrogen lines at 5666.630, 5676.020, 5679.560 and 5686.210 {\AA} and oxygen at 4661.632 {\AA}, respectively. These lines were selected because of the smallest contributions of blends (cf. Fig.\,\ref{fig2} where the N II lines were plotted). The LH models were calculated with the C II atom model, which includes 17 individual levels and 5 superlevels. This atom model does not have the sufficient number of individual levels to obtain correct values of equivalent widths of C II lines located in the visual spectral range. Therefore, we calculated the new non-LTE atmosphere models (T1 - T4 in Table\,\ref{table2}) with the updated C II atom model, which now includes 34 levels individually and 5 superlevels. The predicted ratios of the equivalent widths shown as triangles in Fig.\,\ref{fig3} were calculated for selected stellar parameters assuming GS98 \citep{1998SSRv...85..161G} and AGSS09 \citep{2009ARA&A..47..481A} chemical compositions (cf. Table\,\ref{table2}). The models shown as red diamonds and green dots in Fig.\,\ref{fig3} were calculated for increased abundances of nitrogen (red diamonds) and increased abundances of both helium and nitrogen (green dots). The observations are shown with pluses for three spectra of HD\,163899. As one can see, only models with an increased abundance of nitrogen are able to fit the observations. This conclusion does not depend on the other parameters of the models.

The value of $\mathscr{L}^{'} = 3.85 \pm 0.1$ dex results from the best fit procedure mentioned above. This solution has also well established values of $T_\mathrm{eff} = 23000 \pm 1000$ K, $\log g = 3.0 \pm 0.15$ and $V_\mathrm{rot} \sin i = 65 \pm 5$ $\mathrm{km~s^{-1}}$. The accuracy of these parameters was estimated by considering grids of the LH and T atmosphere models for different chemical compositions and  microturbulent velocities assuming that there is no contribution by additional source of radiation in the cores of hydrogen lines. Otherwise, $\log g = 3.0$ should be regarded as a minimum value of the surface gravity. For instance, if an additional light contributes by amount of about $30$ per cent in the cores of hydrogen lines, then $\log g$ increases to about 3.5 dex. This is the main source of uncertainty in our analysis.

In the case of slowly and moderately rotating stars, the parameter $\mathscr{L}^{'}$ can be directly compared with $\mathscr{L} = \log (L / M) - \log (L_\odot / M_\odot)$ constrained from evolutionary models and spectroscopic Hertzsprung-Russel (sHR) diagrams introduced by \citet{2014A&A...564A..52L}. These diagrams with ($\mathscr{L} = \mathscr{L}^{'}$) vs. $T_\mathrm{eff}$ or vs. $\log g$ can be used to derive stellar parameters such as mass, luminosity, age, etc. An additional advantage is that sHR diagrams are not influenced by interstellar extinction and distance to the star.

HD\,163899 is now more luminous and slightly hotter than assumed in the previous papers (eg. Fig.\,4 in \citealt{2006ApJ...650.1111S}). As a result, it is now more massive. The older spectral classification of HD\,163899 was B2 Ib/II, but the value of $\log g = 3.0$, derived in this paper corresponds better to B2 II.

\section{Evolutionary models} \label{models}

The evolutionary models were computed by means of \texttt{MESA} code \citep{2011ApJS..192....3P,2013ApJS..208....4P,2015ApJS..220...15P}. We assumed the solar chemical composition ($Z=0.0142$, $X=0.7155$, AGSS09) and the OPAL opacity tables \citep{1996ApJ...464..943I}. We used the Ledoux criterion for the convective instability with the value of the mixing-length parameter of $\alpha_\mathrm{MLT} = 1.82$ (calibration of \citealt{2016ApJ...823..102C}). The exponential formula of \citet{2000A&A...360..952H} for overshooting from the convective regions has been used, i.e.:
\begin{equation} \label{eq_overshooting}
  \label{overshooting}
  D_\mathrm{OV} = D_\mathrm{conv}\exp(-\frac{2z}{fH_P}),
\end{equation}
\noindent where $D_\mathrm{conv}$ is the diffusion coefficient derived from Mixing Length Theory (MLT) at a user-defined location in the convective zone ($f_0 H_P$ off the boundary of a convective zone), $H_P$ is the pressure scale height at that location, $z$ is the distance in the radiative layer away from that location and $f$ is an adjustable parameter. In regions stable according to the Ledoux criterion for convection but unstable according to the Schwarzschild criterion we use a semiconvective mixing with the scheme of \citet{1983A&A...126..207L,1985A&A...145..179L} and the standard value of the efficiency parameter is $\alpha_\mathrm{SC} = 0.01$.

The rotation in \texttt{MESA} is implemented in the framework of the shellular approximation, which means that the angular velocity, $\Omega$, is constant over isobars (e.g. \citealt{1997A&A...321..465M}). Mixing of chemical elements and transport of angular momentum are treated in the diffusion approximation \citep{2000ApJ...528..368H,1978ApJ...220..279E,1989ApJ...338..424P}. This approach is consistent with implementation of all others mixing processes in \texttt{MESA} such as convection, semiconvection, overshooting etc. Different rotational mixing mechanisms were taken into account: dynamical shear instability, Solberg-H{\o}iland instability, secular shear instability, Eddington-Sweet circulation and Goldreich-Schubert-Fricke instability. We neglected all effects of magnetic field hence the effects of Spruit-Tayler dynamo were not taken into account \citep{2002A&A...381..923S}. The adopted efficiency of rotational mixing is based on the calibration of \citet{2011A&A...530A.115B}.

We calculated models with initial values of rotation in a range of 20$\%$ - 50$\%$ of the critical velocity on the zero age main sequence (ZAMS). The critical rotational rate is given by $\Omega_\mathrm{crit}^2 = GM/R_\mathrm{eq}^3$ within the Roche model, where $R_\mathrm{eq}$ is the equatorial radius. These angular velocities at ZAMS correspond to the rotational velocities of about $V_\mathrm{rot} = 140$ - $350$ km/s for all models in the considered range of masses. The rotation rate evolves with time due to changes of the internal structure and angular momentum transport. Mass loss is parameterized using the prescription of \citet{2001A&A...369..574V}, which is often used for massive stellar models.

\subsection{The evolutionary stage of HD\,163899} \label{evolutionary_stage}

\begin{figure*}
\begin{center}
\includegraphics[clip,width=88mm,angle=0]{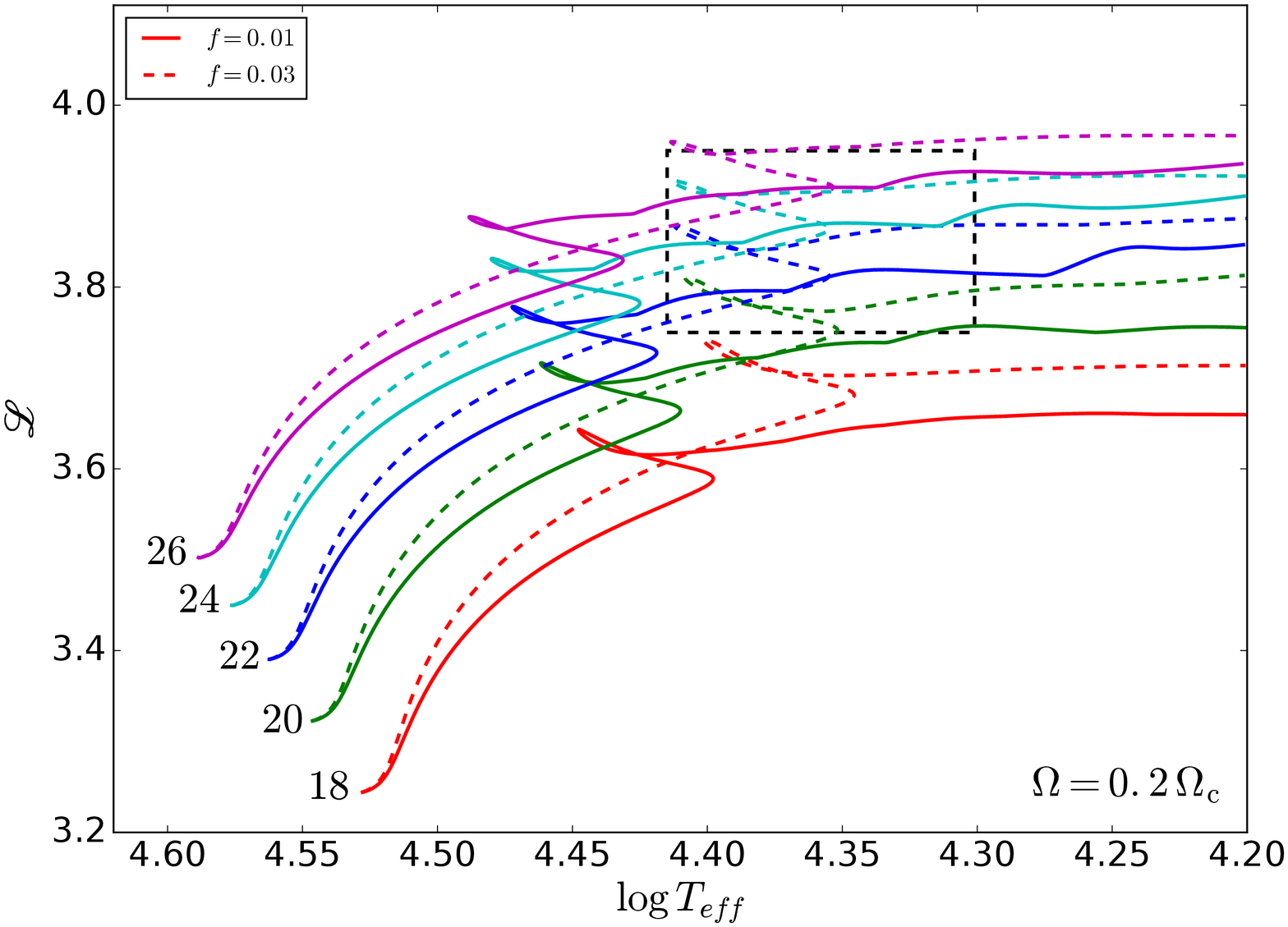}
\includegraphics[clip,width=88mm,angle=0]{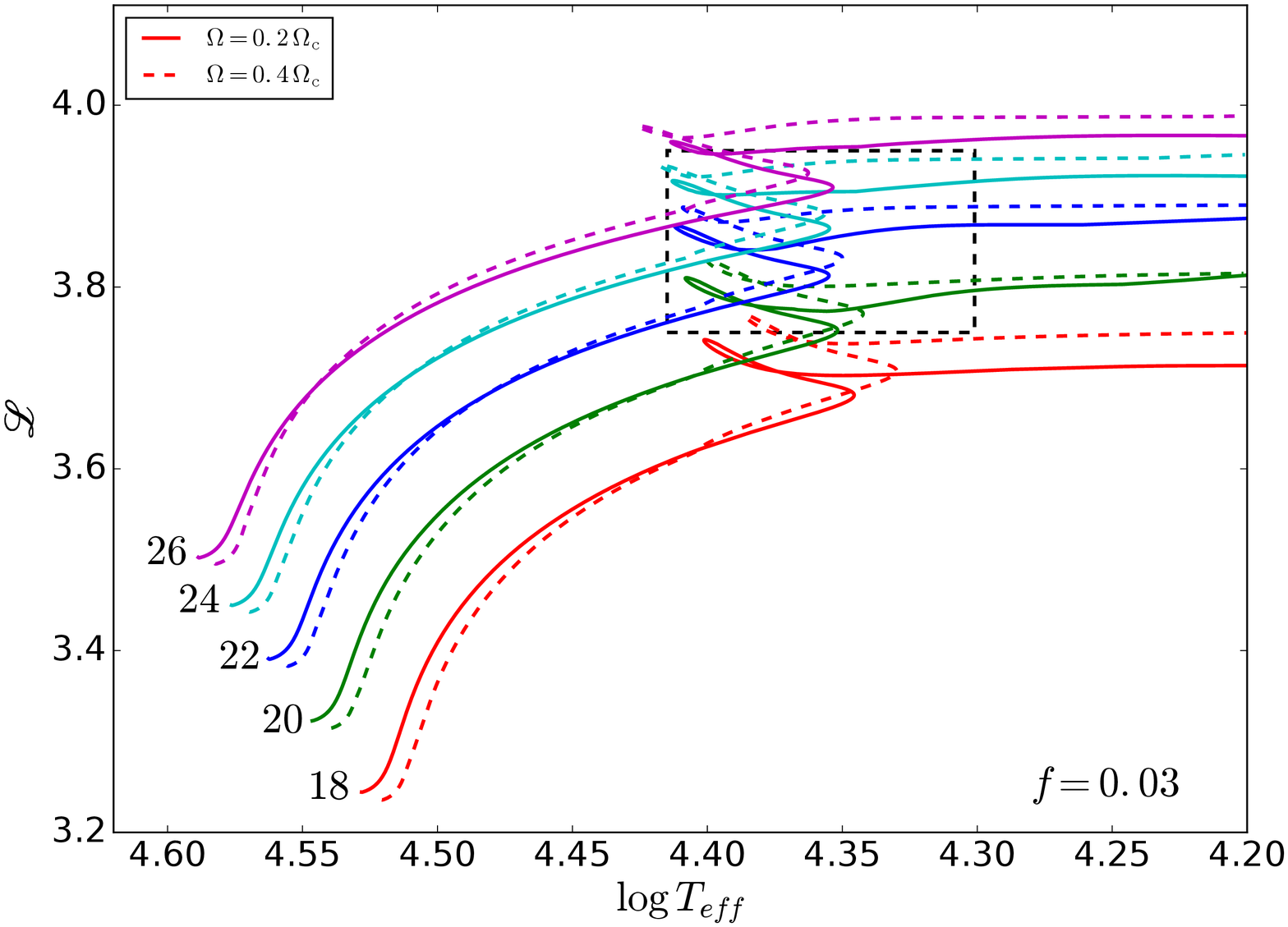}
  \caption{pectroscopic HR diagrams ($\mathscr{L}$ vs. $\log T_\mathrm{eff}$) for models with the initial masses in the range 18 - 26 $M_\odot$ and solar metallicity, $Z = 0.0142$. In the left panel, the effect of convective overshooting is presented. Solid and dashed lines show evolutionary tracks with the overshooting parameter $f=0.01$ and $f=0.03$, respectively. In the right panel, the effect of rotation is shown. Models with rotation rates $\Omega = 0.2~\Omega_\mathrm{crit}$ and $\Omega = 0.4~\Omega_\mathrm{crit}$ are depicted with solid and dashed lines, respectively. The angular velocities of $0.2\Omega_\mathrm{crit}$ and $0.4\Omega_\mathrm{crit}$ at ZAMS corresponds to rotational velocities $V_\mathrm{rot} \simeq 140$ and $280$ km/s for all stars in the presented range of masses. The rectangle shows error box for HD\,163899.}
\label{fig4}
\end{center}
\end{figure*}

\begin{figure}
\begin{center}
\includegraphics[clip,width=87mm,angle=0]{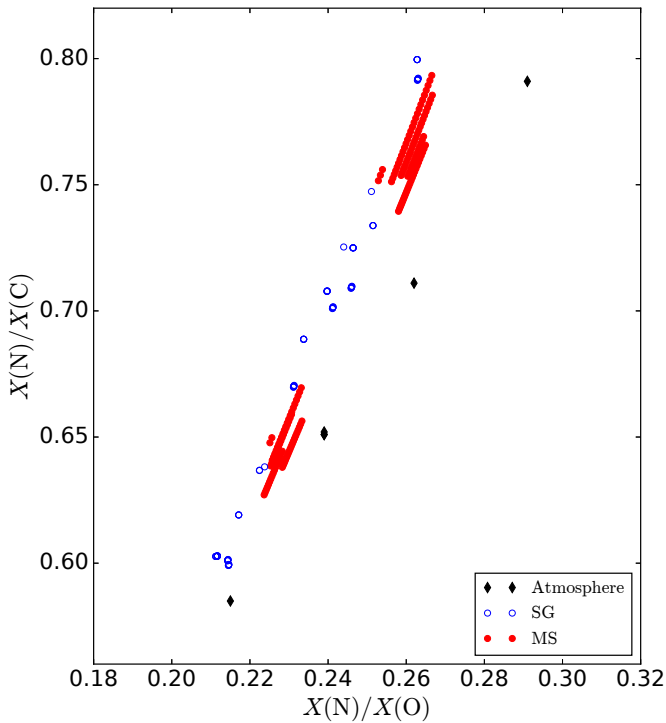}
  \caption{Ratios of mass abundances of $^{14}$N and $^{12}$C, $X(\mathrm{N})/X(\mathrm{C})$, vs. ratios of mass abundances of $^{14}$N and $^{16}$O, $X(\mathrm{N})/X(\mathrm{O})$, for values derived from atmosphere models with nitrogen enrichment (black diamonds; Sec.\,\ref{hd163899}) and for evolutionary models in the error box of HD\,163899 on the main-sequence (red dots) and supergiant phase (blue open circles). The selection of evolutionary models is discussed in the Sec.\,\ref{models}. This figure compares to the area in Fig.\,\ref{fig3} where T3 and T4 models as well as observational values for HD\,163899 are located.}
\label{fig5}
\end{center}
\end{figure}

New determinations of the basic parameters of HD\,163899 and the projected value of the rotational velocity, $V_\mathrm{rot}\sin i$, once again opened up the possibility that the star can be an evolved main sequence star. In Fig.\,\ref{fig4}, we showed the luminosity to mass ratio, $\mathscr{L}$, vs. effective temperature, $\log T_\mathrm{eff}$, for models with initial masses of 18, 20, 22, 24 and 26 $M_\odot$. The chosen masses are suitable to reproduce the observed parameters of the star. The presented evolutionary tracks cover the phase from ZAMS up to the point where $\log T_\mathrm{eff}=4.2$. In the left panel, we show the effect of the convective overshooting parameter, $f$, for the fixed value of the rotation rate at ZAMS. We set 0.2$\Omega_{\rm crit}$, which corresponds to the rotational velocity $V_\mathrm{rot} \simeq 140$ km/s for all masses presented here. In the right panel, the effect of rotation is showed for $f=0.03$. The width of the main sequence is mostly determined by the overshooting from the hydrogen core (the left panel of Fig.\,\ref{fig4}), whereas the faster rotation shifts the evolutionary track up to the higher values of luminosity (the right panel of Fig.\,\ref{fig4}). As we will see in the next subsection, the moderate rotation has a significant impact on the internal structure. In turn, the internal structure determines the pulsational properties of a star which we will explore hereafter.

With the new parameters, the star is more luminous than it was believed before and slightly higher values of the effective temperature are preferred as well. These new values have two major consequences. Firstly, the required initial mass for models of HD\,163899 should now be at least as high as about 20 $M_\odot$. For the previous estimates of stellar parameters, we found that models with masses of around 15-16 $M_\odot$ are the most appropriate \citep{2015MNRAS.447.2378O}. Secondly, main-sequence models are now contained in the error box if sufficiently high values of the convective overshooting parameter are used ($f\ge 0.02$, cf. the left panel of Fig.\,\ref{fig4}).

The effect of the rotationally induced mixing can be seen in stellar atmospheres.
In Section\,\ref{hd163899}, we found an overabundance of nitrogen from the spectral analysis of HD\,163899. To confront this result with theoretical predictions, in Fig.\,\ref{fig5} we present the ratio of mass abundances of $^{14}$N and $^{12}$C, $X(\mathrm{N})/X(\mathrm{C})$, vs. ratio of mass abundances of $^{14}$N and $^{16}$O, $X(\mathrm{N})/X(\mathrm{O})$ for values derived from atmosphere models with nitrogen enrichment (black diamonds; Sec.\,\ref{hd163899}) and for evolutionary models in the error box of HD\,163899 on the main-sequence (red dots) and supergiant phase (blue open circles). Fig.\,\ref{fig5} can be compared to the area in Fig.\,\ref{fig3} where T3 and T4 models as well as observational values of equivalent widths ratios for HD\,163899 are located. We did not show atmosphere models without nitrogen enrichment in Fig.\,\ref{fig5} because they are not suitable for the studied star. In the rotating models the nitrogen abundance is increasing during most of the main sequence evolution and during the phase of second contraction, whereas for the non-rotating models (not shown) the abundance of $^{14}$N is constant during the studied phases of the evolution. Thus, it seems that the nitrogen overabundance of HD\,163899 can be reproduced only by rotating models.

The parameters of studied stellar models are chosen to reproduce the abundance ratios of $X(\mathrm{N})/X(\mathrm{C})$ and $X(\mathrm{N})/X(\mathrm{O})$ suitable for the atmosphere models of HD\,163899 ($X(\mathrm{N})/X(\mathrm{C}) \in [0.58, 0.80]$, $X(\mathrm{N})/X(\mathrm{O}) \in [0.21, 0.30]$) and the determined parameters of the star within their error boxes (Section\,\ref{hd163899}). For comparison, $X(\mathrm{N})/X(\mathrm{C}) = 0.29$ and $X(\mathrm{N})/X(\mathrm{O})=0.12$ for AGSS09 mixture. The full selection of models that fulfil these criteria is presented in Fig.\,\ref{fig5}. We used the observed values to constraint the rotational velocity for models with different combinations of initial mass and overshooting. Generally, with higher initial mass, the rotation rate required to obtain the observed nitrogen overabundance is lower. The overshooting parameter also influences the surface abundances because it prolongs the MS phase, during which the rotational mixing is efficient. For all selected models the initial rotation rate is in range of $0.26-0.30\Omega_\mathrm{crit}$. The linear increase of the surface nitrogen abundance is clearly visibile in the Fig.\,\ref{fig5}: the MS models are divided into two separate groups, one with lower values of $X(\mathrm{N})/X(\mathrm{C})$ and $X(\mathrm{N})/X(\mathrm{O})$ and one with higher values. The first group corresponds to the models with $\Omega = 0.28~\Omega_\mathrm{crit}$ and the second to $\Omega = 0.30~\Omega_\mathrm{crit}$. This bifurcation is not visible for the supergiant models because of the additional effect of overshooting. On the MS almost all models in the error box have $f=0.03$, whereas supergiants have $f=0.01$ or $0.02$. That leads to continous visible distribution of points despite the same step in $\Omega$ in the studied grid ($\Delta\Omega = 0.02$).

\subsection{Internal structure} \label{internal_structure}

\begin{figure*}
\begin{center}
 \includegraphics[clip,width=180mm,angle=0]{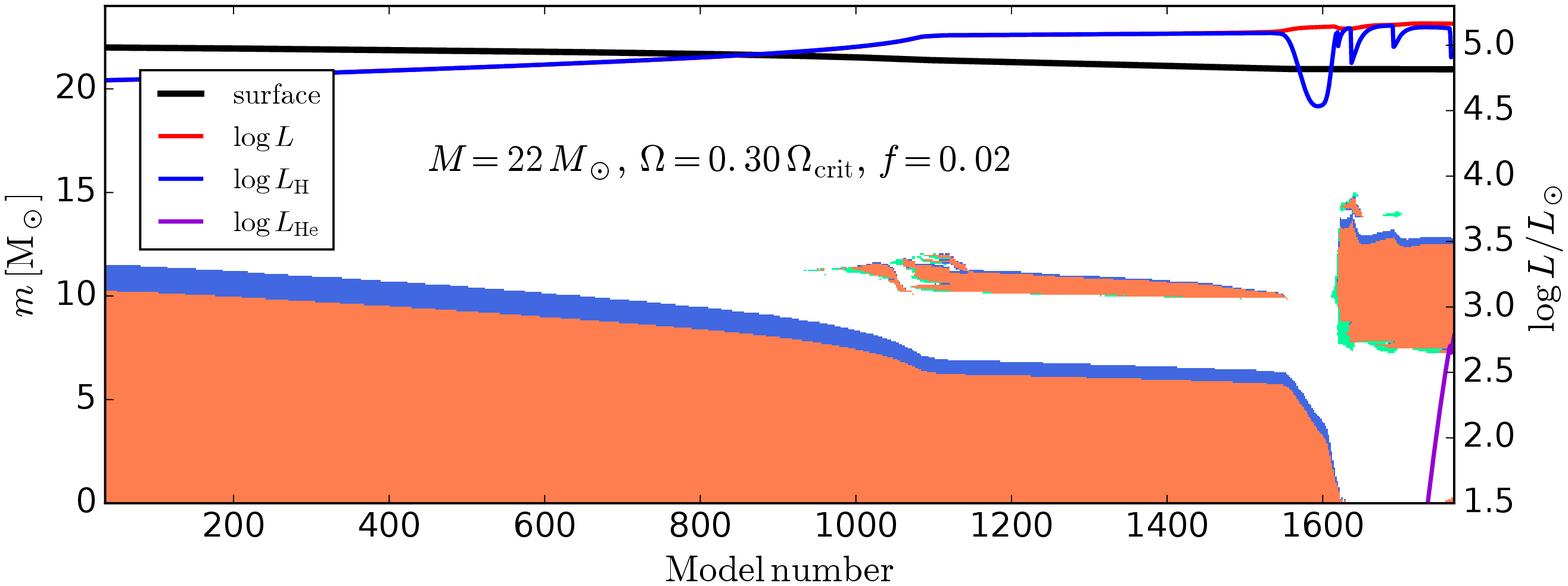}
 \includegraphics[clip,width=180mm,angle=0]{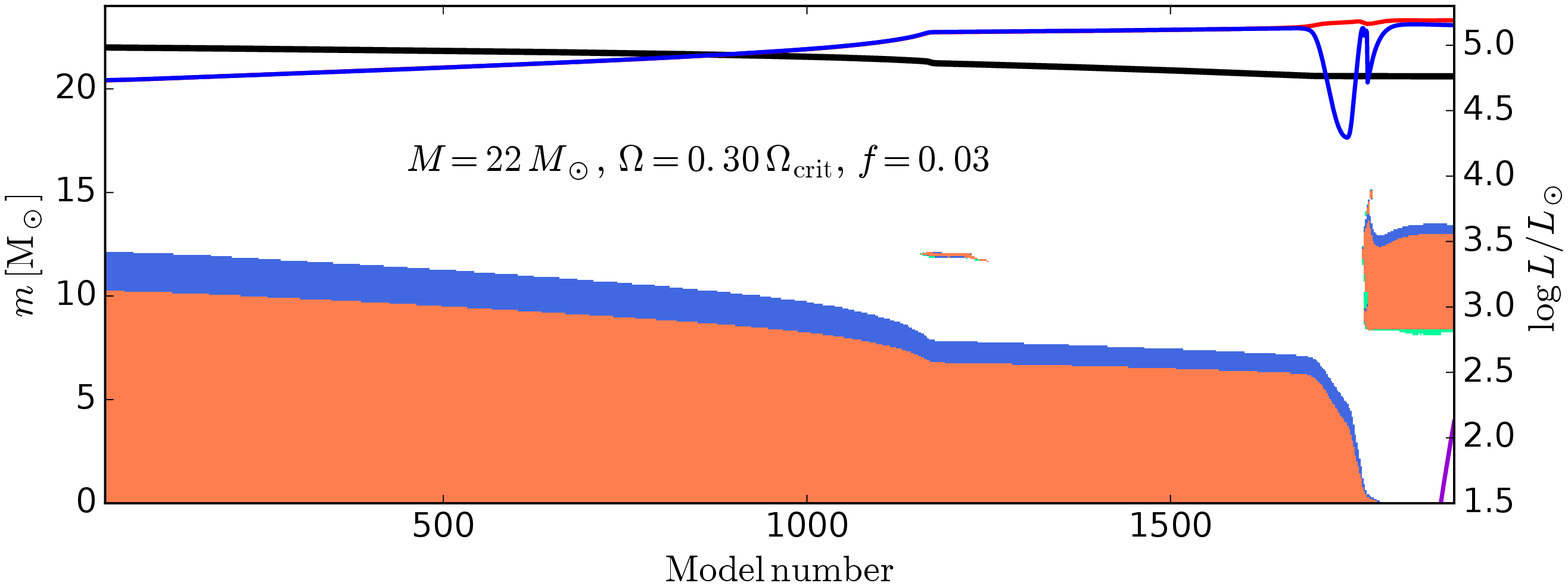}
   \caption{Kippenhahn diagrams for two models with the initial masses of 22 $M_\odot$ and the convective overshooting parameter $f=0.02$ (top panel) and $f=0.03$ (bottom panel). In both cases the value of rotation rate at ZAMS is $\Omega=0.3~\Omega_\mathrm{crit}$. The structure of a star is showed as a function of the calculated model number. Convective zones, convective overshooting and semiconvective zones are showed with orange, blue and green colours, respectively. The white areas depict radiative zones and the surface of the star is presented with a black line. The total luminosity (red lines) as well as luminosities from hydrogen (blue lines) and helium burning (violet lines) are also showed on the righ-hand Y-axis. The presented evolution covers time from ZAMS to the point where $\log T_\mathrm{eff}=4.0$.}
\label{fig6}
\end{center}
\end{figure*}

Modelling massive stars is a difficult task due to a number of uncertainties regarding mixing mechanisms and the sensitivity of global stellar parameters such as $\log T_\mathrm{eff}$ or $\log L$ to the internal structure. The mixing parameters control the size and efficiency of convective and semiconvective zones, determine ranges of convective overshooting and occurrence of zones with rotationally induced instabilities. These different mixing layers are crucial for the transport of energy, chemical elements and angular momentum and through their influence on the available amounts of nuclear fuel, the rates of energy generation from thermonuclear reactions, total flux of the star and other important components of the model. The recent examples of massive star modelling can be found, e.g., in \citet{2011A&A...530A.115B}, \citet{2012A&A...537A.146E}, \citet{2016ApJ...823..102C} and \citet{2013A&A...558A.103G} for very low metallicity.

Changes of the internal structure of a star during its evolution can be easily followed using the Kippenhahn diagrams. In Fig.\,\ref{fig6}, we show such diagrams for two models with the initial masses of 22 $M_\odot$ and the convective overshooting parameter $f=0.02$ (top panel) and $f=0.03$ (bottom panel). In both cases the rotation rate at ZAMS is $30\%$ of $\Omega_\mathrm{crit}$. The structure of a star is showed as a function of a calculated model number and the plots cover the evolution from ZAMS to the point where $\log T_\mathrm{eff}=4.0$ after the overall contraction. Convective zones, convective overshooting and semiconvective zones are showed with orange, blue and green colours, respectively. The radiative zones are depicted as white areas. The total luminosity (red line) as well as luminosities from hydrogen and helium burning (grey and yellow lines, respectively) are also shown. The end of the MS is around model number 1400 ($\log T_\mathrm{eff} \simeq 4.39$) for the case shown in the top panel ($M = 22 M_\odot$, $f=0.02$) and around model number 1660 ($\log T_\mathrm{eff} \simeq 4.35$) for the bottom panel ($M = 22 M_\odot$, $f=0.03$).

The effects of overshooting and rotation on stellar structure are very complex. When the rotation is slow or neglected and none or small overshooting is assumed, 
thin embedded semiconvective layers may occur in convective zones at some evolutionary stages. It is common in \texttt{MESA} models with masses higher than 18 $M_\odot$ but it may also be visible in less massive models. The properties of semiconvective layers have been discussed by, for example, \citet{1991A&A...252..669L}, \citet{1993SSRv...66..409M}, \citet{2009CoAst.158...79C} or \citet{2016ApJ...817...54M}. Such a non-smooth structure has rather small effect on the shape of evolutionary tracks or the energy output of models in the considered range of effective temperatures ($\log\/T_\mathrm{eff} > 4.0$), but strongly modifies the chemical composition gradient, $\nabla_\mu$. As a  consequence, such features greatly affect the Brunt-V\"{a}is\"{a}l\"{a} frequency. High order gravity modes, which are common in evolved massive stellar models (cf. Section \,\ref{pulsations}), are very sensitive to the fine structure of a star, especially around the intermediate convective zone (ICZ, e.g. \citealt{2009A&A...498..273G}). The effect of layered semiconvection on properties and period spacing of high-order g-modes has been studied by \citet{2008MNRAS.386.1487M} and more recently by \citet{2015MNRAS.452.2700B}. Both papers relied on the adiabatic approximation. For high-order g-modes, discontinuities in the ICZ may lead to the emergence of additional acoustic cavities. Therefore, when one computes pulsational instability, the resulting oscillation spectra might contain fictitious unstable modes.

Models with moderate rotation rate, $\Omega \simeq 0.3\/\Omega_\mathrm{crit}$, which we found suitable for the analysed star, are less prone to the occurence of layered semiconvection. This is due to more efficient rotationally induced mixing. Examination of Fig.\,\ref{fig6} confirms this fact: the ICZ is free of embedded semiconvection in both presented cases. The values of the convective overshooting parameter used in the considered models are in range $f=0.01-0.03$. The convective overshooting has two important effects on the studied models. Firstly, higher values of $f$ lead to a wider main sequence (Sec.\,\ref{evolutionary_stage}, Fig.\,\ref{fig4}). Secondly, with the more efficient overshooting from the hydrogen core, the ICZ during the MS evolution is either thinner or it does not appear at all. This effect may be evaluated by comparing the top ($f=0.02$) and bottom ($f=0.03$) panels of Fig.\,\ref{fig6}. These conclusions are true for all studied masses.

\section{Interpretation of the observed frequencies of HD 163899}

The analysis of the MOST light curve of HD 163899 by \citet{2006ApJ...650.1111S} revealed 48 frequency peaks in the range [0.02, 2.85] d$^{-1}$.
The observations spanned 37 days thus the Rayleigh resolution limit, $1/T$, is about 0.03 d$^{-1}$. Therefore,
the frequency $\nu=0.021$ should not be considered to be intrinsic to the star. Moreover, some frequencies listed in \citet{2006ApJ...650.1111S} do not fulfill
the signal-to-noise ratio requirement, $S/N$. The commonly adopted criterion for the frequency detection is that $S/N$ should be around 4 \citep{1993A&A...271..482B}. Besides, the photometric amplitudes increase towards the low frequencies which can be interpreted as the "red noise" excess  (e.g., \citealt{2011A&A...533A...4B}). The fit of the power law shape, $\nu^a$, gives the value $\chi^2=0.23$ with the power $a=-0.34(4)$.
Thus, some of these frequencies can have a stochastic origin. Another possibility is that HD 163899 is a binary and the pulsations come from a companion. If the latter has a mass less than about 7 $M_\odot$ and is a fast rotator then the oscillation spectrum can be easily explained by the pulsational models (e.g., \citealt{2005MNRAS.364..573T,2007MNRAS.374..248D}).

\subsection{Pulsational models} \label{pulsations}

In this section, we allow the hypothesis that the observed frequencies are associated with the heat-driven pulsations which come from HD 163899 itself.
For this purpose, we accepted frequencies which have $S/N \ge 3.9$ and excluded the lowest frequency around 0.02 d$^{-1}$.
Thus, we considered 44 frequencies out of 48 in total (see the top panel of Fig.\,8).

\begin{figure}
\begin{center}
\includegraphics[clip,width=87mm,angle=0]{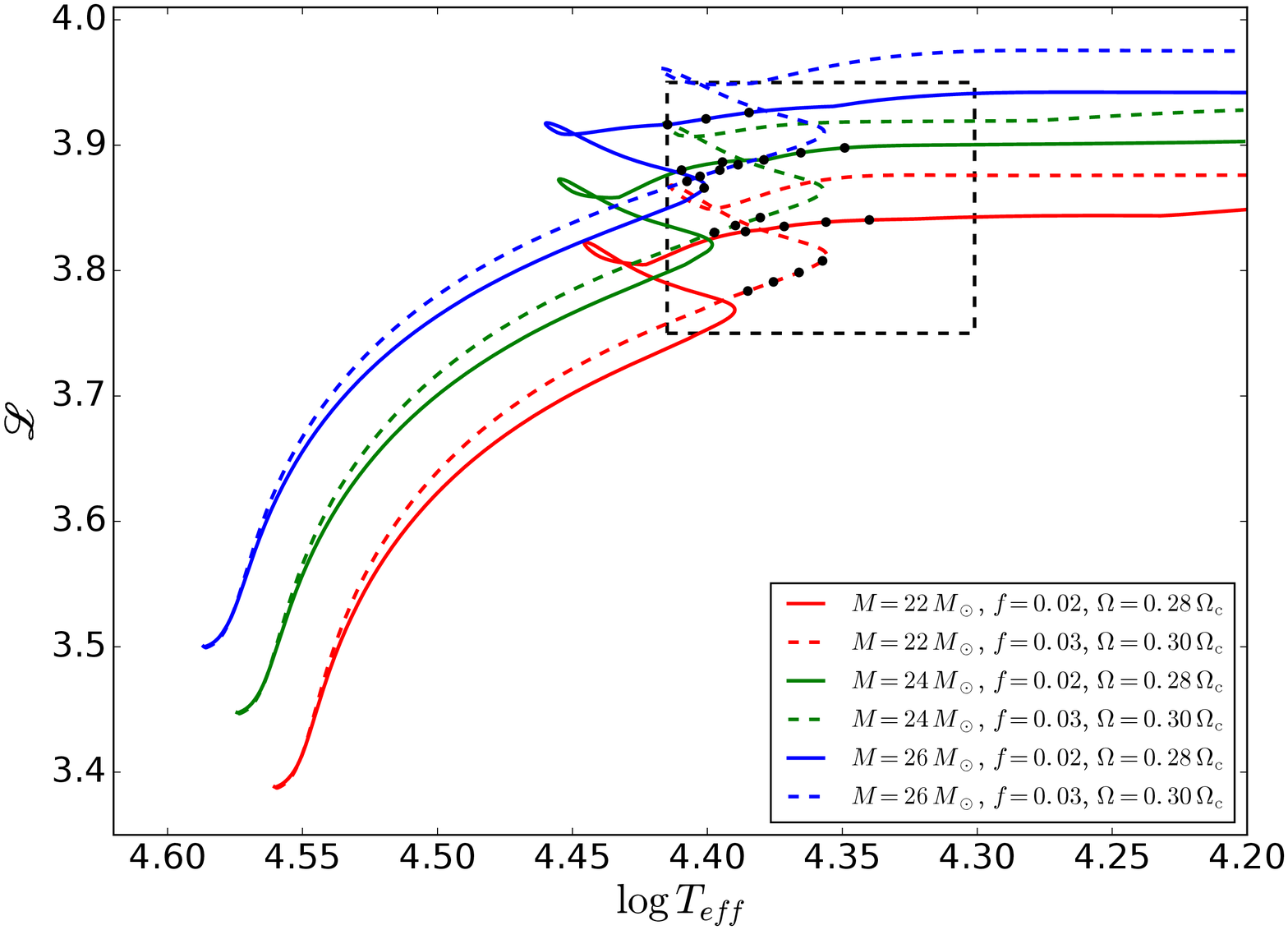}
  \caption{Evolutionary tracks in the spectroscopic HR diagram, $\mathscr{L}$ vs. $\log T_\mathrm{eff}$, for selected models contained within the error box of HD\,163899 (black dots) with initial masses of 22, 24 and 26 $M_\odot$, overshooting $f=0.02$ and $0.03$ and initial rotation rates at ZAMS of 0.28 and 0.30 $\Omega_\mathrm{crit}$.}
\label{fig7}
\end{center}
\end{figure}

\begin{figure*}
\begin{center}
\includegraphics[clip,width=180mm,angle=0]{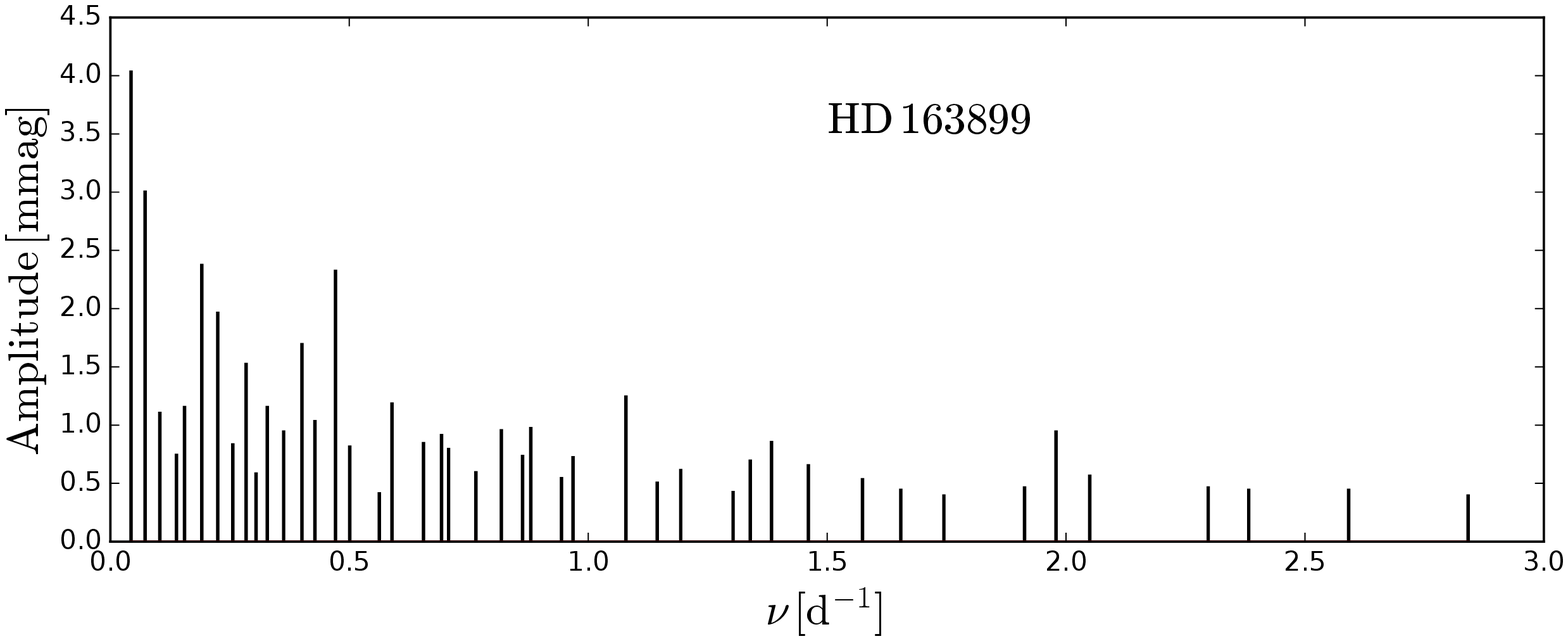}
\includegraphics[clip,width=180mm,angle=0]{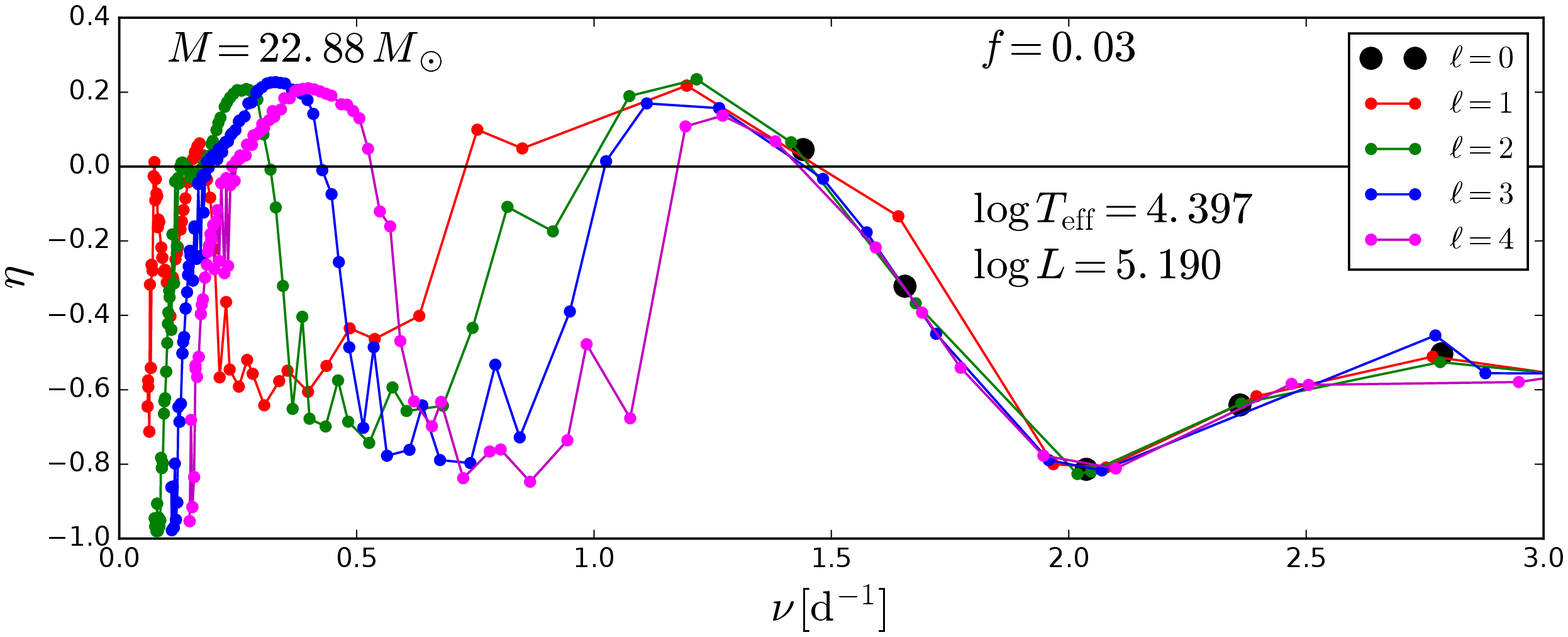}
\includegraphics[clip,width=180mm,angle=0]{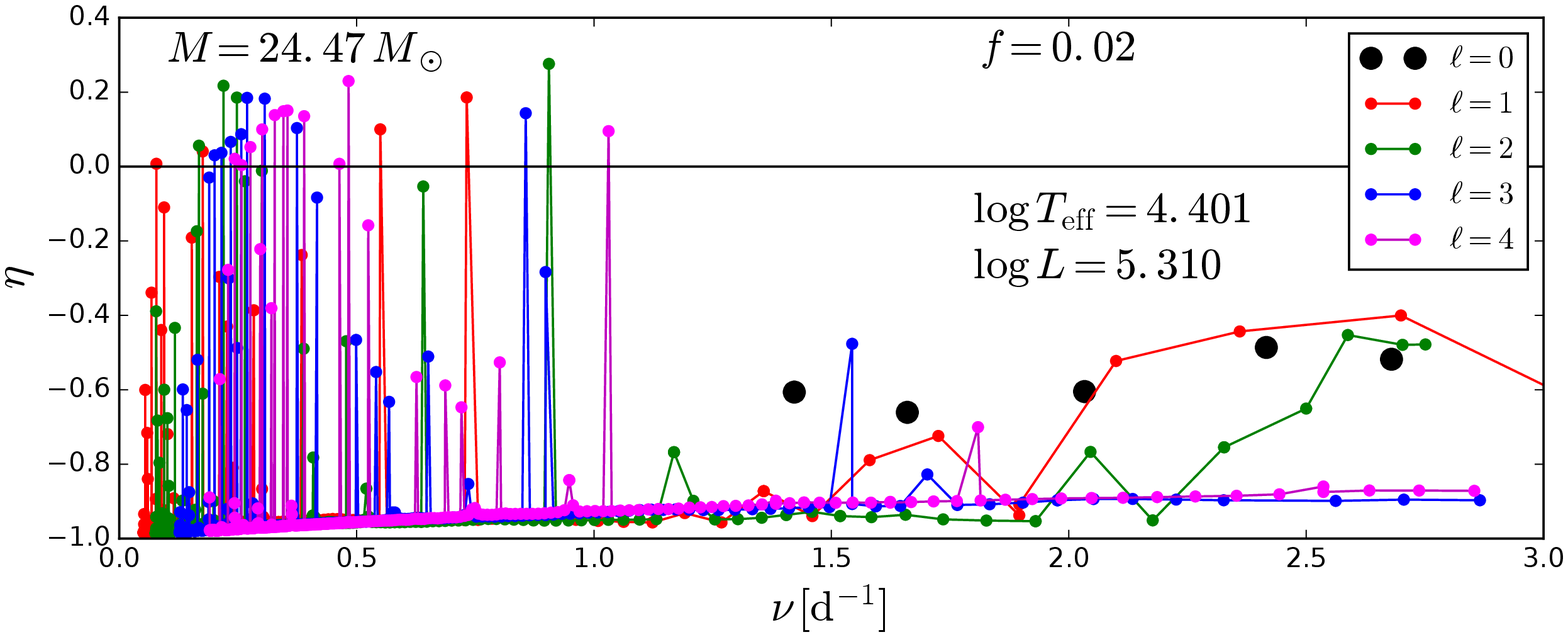}
  \caption{Comparison of the observed frequency spectrum of HD\,163899 (the top panel, c.f. Table 1 in \citealt{2006ApJ...650.1111S}) and theoretical frequencies calculated for a main-sequence model with a mass of 22.88 $M_\odot$ (the centre panel) and a supergiant model with a mass of 24.47 $M_\odot$ (the bottom panel). Modes with harmonic spherical degree up to $\ell = 4$ are shown.}
\label{fig8}
\end{center}
\end{figure*}

We computed pulsations for massive stellar models satisfying the N/C criterion discussed in the previous sections.
In Fig.\,\ref{fig7}, we put the representative models on the $\mathscr{L}$ vs. $\log T_\mathrm{eff}$ diagram together with the evolutionary tracks. The presented models have the initial masses $M_0= 22,~24$ and 26 $M_\odot$, overshooting parameter $f=0.02$ and $0.03$ and the initial rotation rates at ZAMS of 0.28 and 0.30 $\Omega_\mathrm{crit}$.
The corresponding values of the rotational frequency, $\nu_{\rm rot}$, of the models are of the order of 0.06 d$^{-1}$ thus the effects of rotation on stellar pulsation can be safely neglected because the spin parameter, $s=2 \nu_{\rm rot}/\nu_{\rm puls}$, is less than 1 for all but three frequencies.

Interestingly, the three lowest frequencies, 0.0431, 0.0726 and 0.1035 d$^{-1}$, show spacing of about 0.03 d$^{-1}$, which is a half of the rotational frequencies of the models. The value of 0.03 d$^{-1}$ is close to the Rayleigh resolution limit, but there is also a possibility that these three peaks are a rotational split triplet with $\ell=1$ because the Ledoux constant for high order g-modes is about 0.5.

The pulsations were computed with the nonadiabatic code of \citet{1977AcA....27...95D} and the results are presented for the two representative models in Fig.\,\ref{fig8}.
The middle and bottom panels show the instability parameter, $\eta$, as a function of the mode frequency, $\nu$, for an evolved main-sequence model with $M=22.88 M_\odot$ ($M_0=24~M_\odot$), $f=0.03$, $\log T_\mathrm{eff}=4.397$, $\log L=5.190$ and a supergiant model with $M=24.47 M_\odot$ ($M_0= 26~M_\odot$), $f=0.02$, $\log T_\mathrm{eff}=4.401$, $\log L=5.310$, respectively.

The instability parameter, $\eta$, measures the net energy gained by a mode during one pulsational cycle and is defined as \citep{1978AJ.....83.1184S}:

\begin{equation}
  \label{eta}
  \eta=\frac{W}{\int\limits_0^R\left|\frac{\mathrm{d}W}{\mathrm{d}r}\right|\mathrm{d}r},
\end{equation}

\noindent where $W$ is the global work integral. This definition gives normalization $\eta \in [-1,+1]$ and the mode is excited if $\eta > 0$.

Here, we considered modes with the harmonic degree up to 4. Although the disc averaging factor, $b_\ell$, drops significantly between $\ell = 2$ and $\ell = 3$,
the geometrical effect in the light variation, $(\ell - 1)(\ell + 2)$ increases with $\ell$ \citep{2006MmSAI..77..113D}. Thus, it is justified to take into account modes higher than $\ell=2$ for interpretation of space data.

The models presented in Fig.\,\ref{fig8} are confined in the error box of HD\,163899 and have the N/C abundance as determined from observations.
As one can see, the two oscillation spectra are dissimilar. In the case of the MS model, there are two distinctive global maxima of $\eta(\nu)$: one related to the high-order g-modes and located at the low frequencies and the second one, at higher frequencies, related to the p and mixed modes.
The fundamental radial mode is unstable in this model and has the frequency of 1.44 d$^{-1}$. The total range of unstable modes is [0.07, 1.44] d$^{-1}$.

The oscillation spectrum of the supergiant model looks quite different. In the low frequency range, below about 1.1 d$^{-1}$, individual pulsational modes become unstable.
The higher frequency modes are all stable, in particular, the radial modes are not excited.
In our previous studies \citep{2013MNRAS.432.3153D}, we obtained for lower-mass supergiant models the consecutive pattern of the local minima and maxima of $\eta$
at low frequencies. This pattern was related to the partial trapping of pulsation modes. For more massive and rotating models, it seems to be not so clearly visible.

The main difference between the main-sequence and supergiant models is the substantially higher number of unstable modes in the MS models.
However, as one can see, none of the models can account for the whole range of the observed frequencies. The noteworthy feature of the observed  spectrum is a rather well
coverage of the frequency range [0.04, 1.75] d$^{-1}$ and a few single peaks above 1.9 d$^{-1}$.
The main problem with the models is the lack of unstable modes with the lowest and highest detected frequencies.
In our next step, we will check whether the disagreement can be solved if the rotational splitting of pulsational frequencies is included .

\subsection{Distribution of pulsational frequencies} \label{frequencies}

For a more reliable comparison of the observed frequencies of HD\,163899 with the theoretical ones,
we calculated the frequency distributions predicted by the pulsational models and confronted them with the observed one.
To this end, we took into account rotational splitting for all the unstable frequency modes ($\eta>0$) according to the formula:

\begin{equation}
  \label{nu_rot}
  \nu=\nu_0+m(1-C_{n\ell})\nu_{\rm rot},
\end{equation}

\noindent where $m$ is the azimuthal order, $C_{n\ell}$ is the Ledoux constant resulting from pulsational models and $\nu_{\rm rot}$ is the rotational frequency.
The value of $\nu_{\rm rot}$, was determined for the rotational velocity, $V_{\rm rot}$, and the radius, $R$, of a given model.
In all cases $\nu_{\rm rot}$ was of the order of 0.06 d$^{-1}$.

From all unstable modes, we selected those which had the theoretical photometric amplitudes in the Johnson $V$ passband higher than the minimum observed amplitudes in the MOST data. The value of this minimum amplitude is $A_{\rm min}=0.039$ mmag.
We computed the photometric amplitudes according to the linear formula of \citet{2002A&A...392..151D}.
This formula contains one pulsational parameter which is not provided in the framework of the linear theory. This is the intrinsic mode amplitude, $\varepsilon$,
defined by the local radial displacement of the photosphere
\begin{equation}
\delta r(R,\theta,\varphi)= R {\rm Re}\{ \varepsilon Y_\ell^m {\rm e}^{-{\rm i}2\pi\nu_{\rm puls} t}\},
\end{equation}

\noindent where $Y_\ell^m$ is spherical harmonic and Re means the real part. Other symbols have their usual meanings.

We estimated the maximum value of $\varepsilon$ taking the highest observed amplitude from the {\it MOST} data, which is $A\approx 4$ mmag at the lowest frequency $\nu=0.0431$ d$^{-1}$, and assuming that it corresponds to the dipole axisymmetric mode.
The value of the inclination angle, $i$, was obtained from the equation $i=\arcsin (65/V_{\rm rot})$.
We got $\varepsilon_{\rm max}$ of about 0.01 independently of the model. 
For each unstable mode, we randomly drew the value of $\varepsilon$ from the range [0, 0.01].
Then, for the obtained inclination angles, we calculated the photometric amplitudes in the Johnson $V$ passband.
We chose this passband because for the considered range of ($T_{\rm eff}, \log g$) the pulsational amplitude in this band is close to the {\it MOST} amplitude.

\begin{figure*}
\begin{center}
\includegraphics[clip,width=88mm,angle=0]{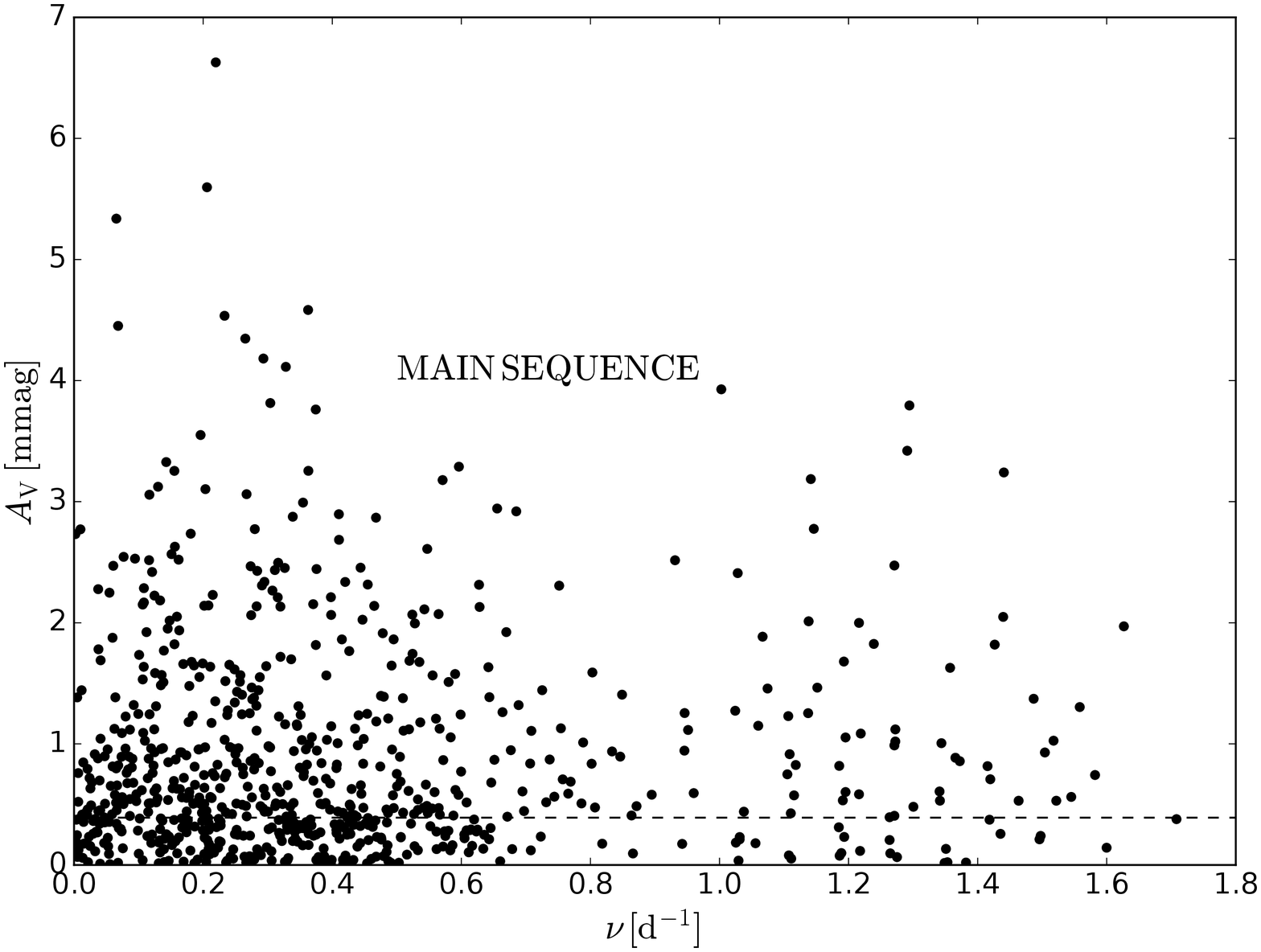}
\includegraphics[clip,width=88mm,angle=0]{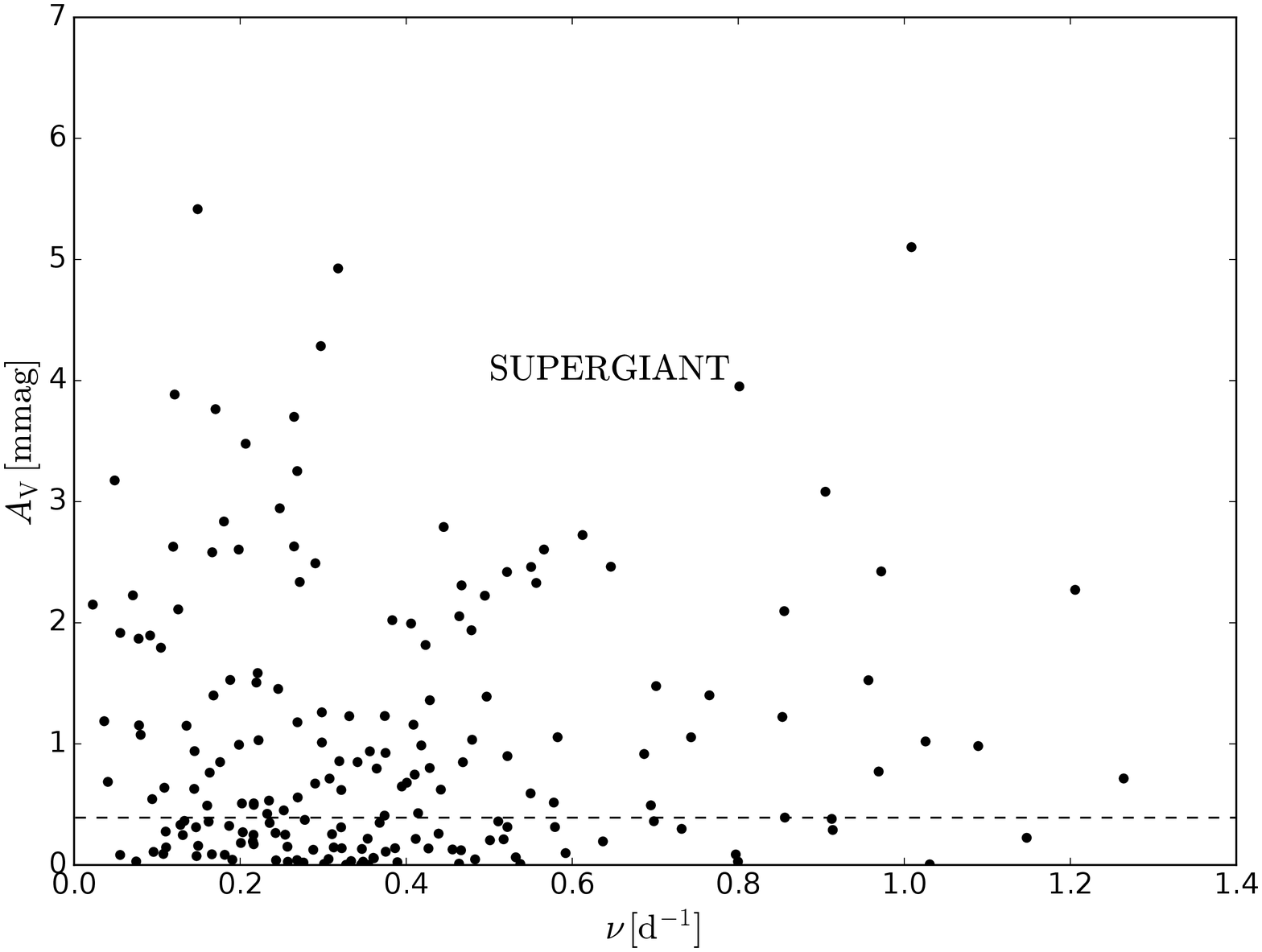}
  \caption{Theoretical photometric amplitudes as a function of the mode frequency for a main-sequence (the left panel) and supergiant (the right panel) models with the initial masses of $24 M_\odot$ and $26 M_\odot$, respectively (the same models as in Fig.\,\ref{fig8}). The horizontal line, located at $A_\mathrm{V} = 0.39$ mmag, depicts the lowest amplitude detected in the MOST data for HD\,163899 with $S/N=3.9$.}
\label{fig9}
\end{center}
\end{figure*}

\begin{figure*}
\begin{center}
\includegraphics[clip,width=88mm,angle=0]{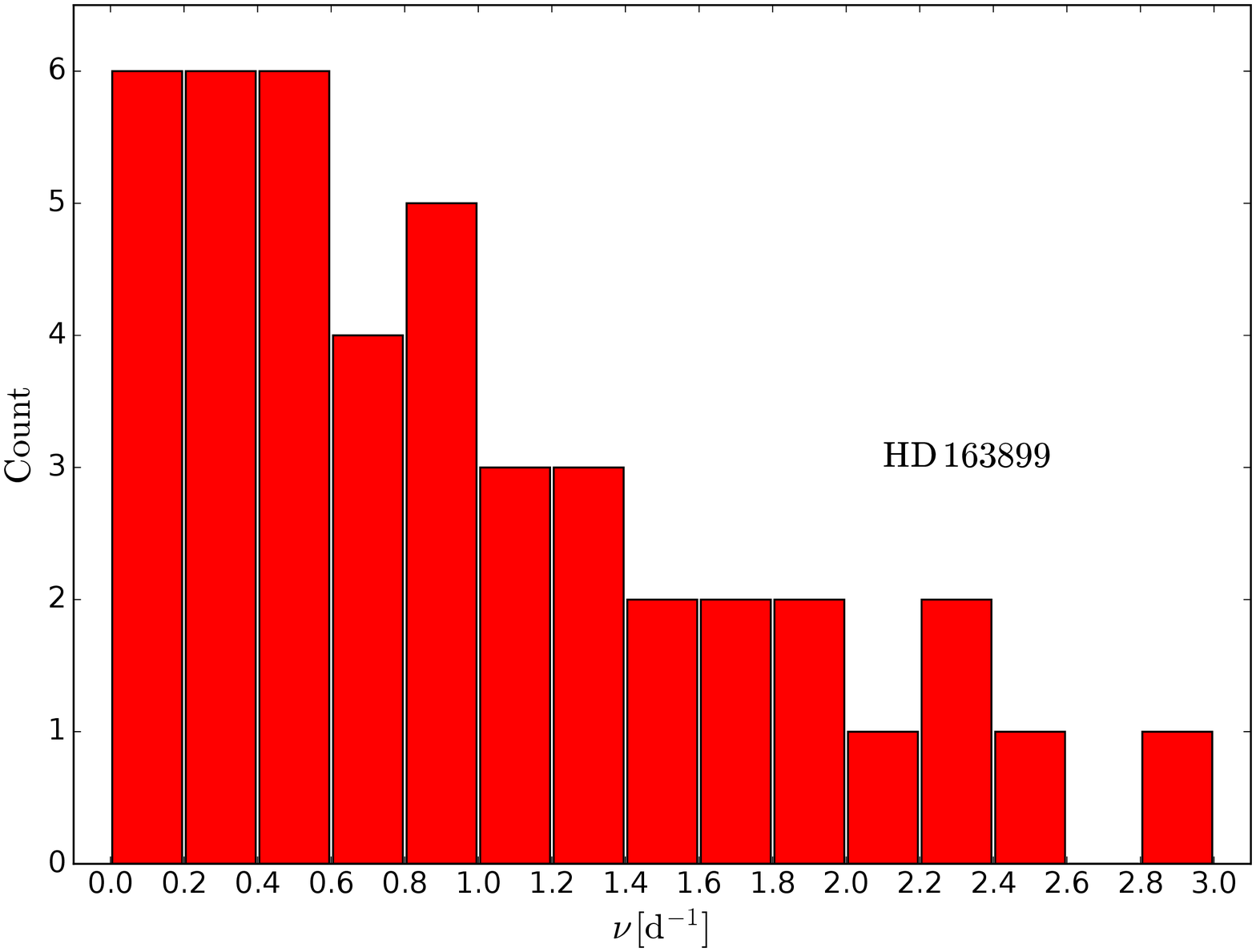}
\includegraphics[clip,width=88mm,angle=0]{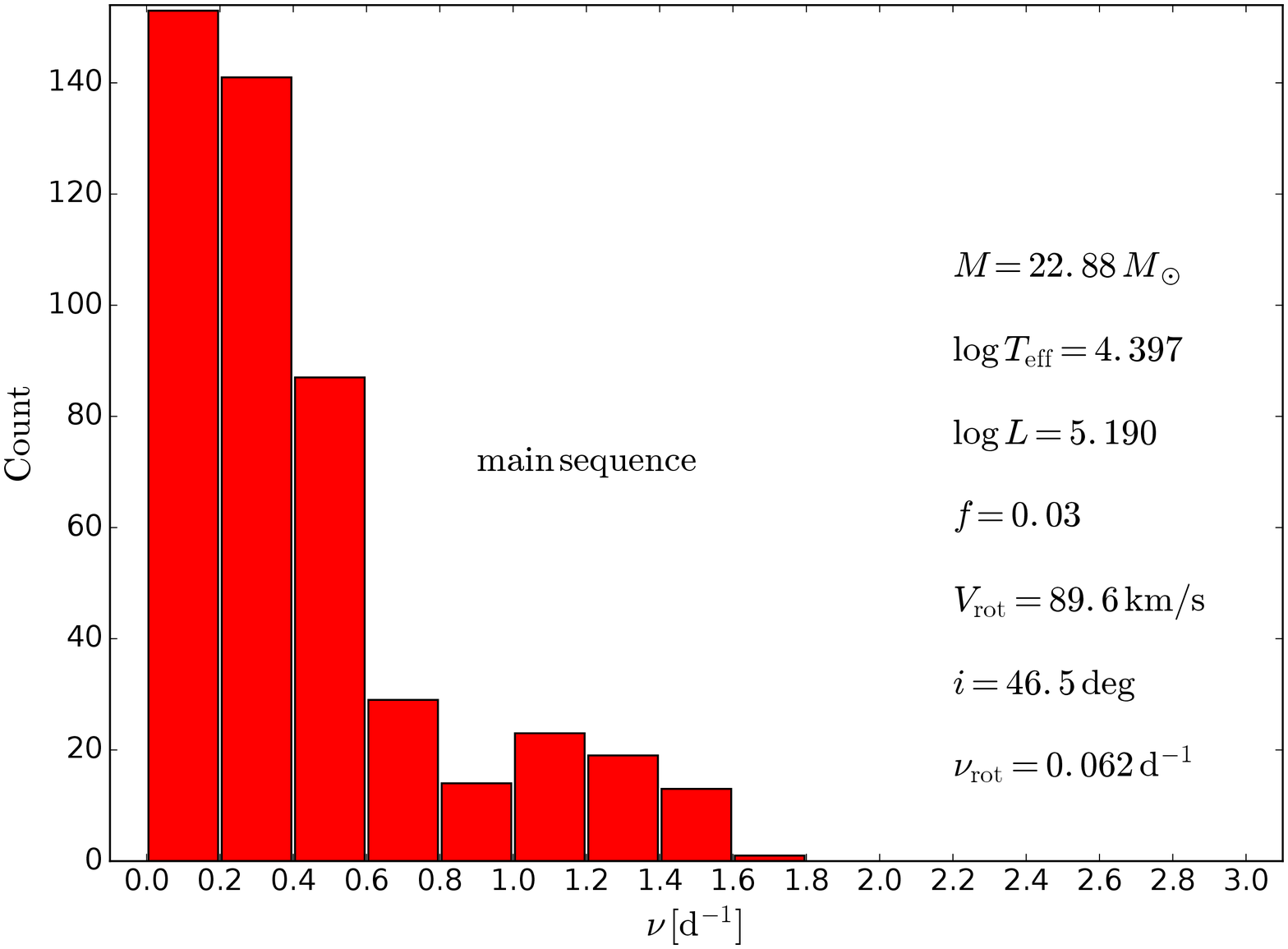}
\includegraphics[clip,width=88mm,angle=0]{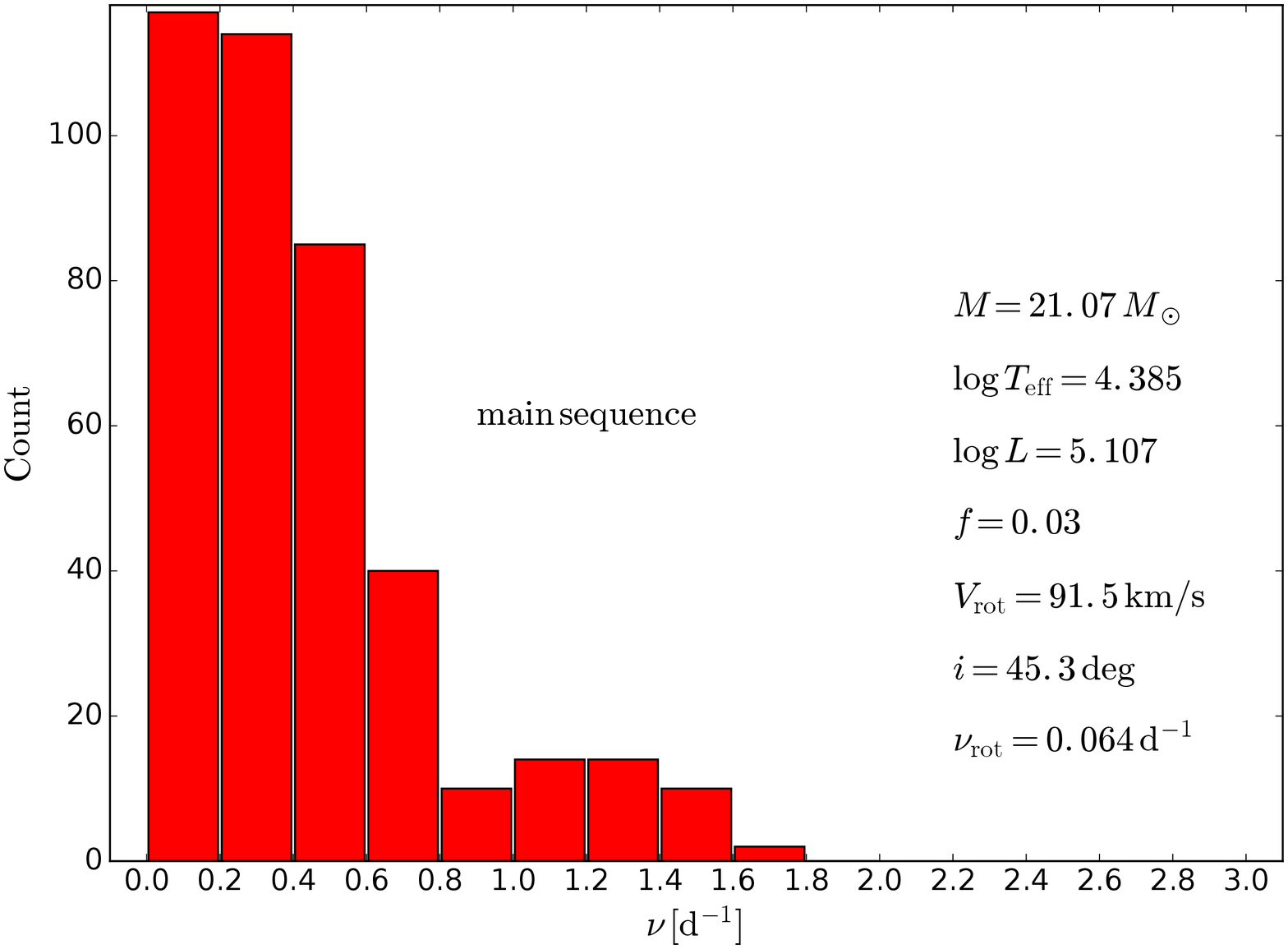}
\includegraphics[clip,width=88mm,angle=0]{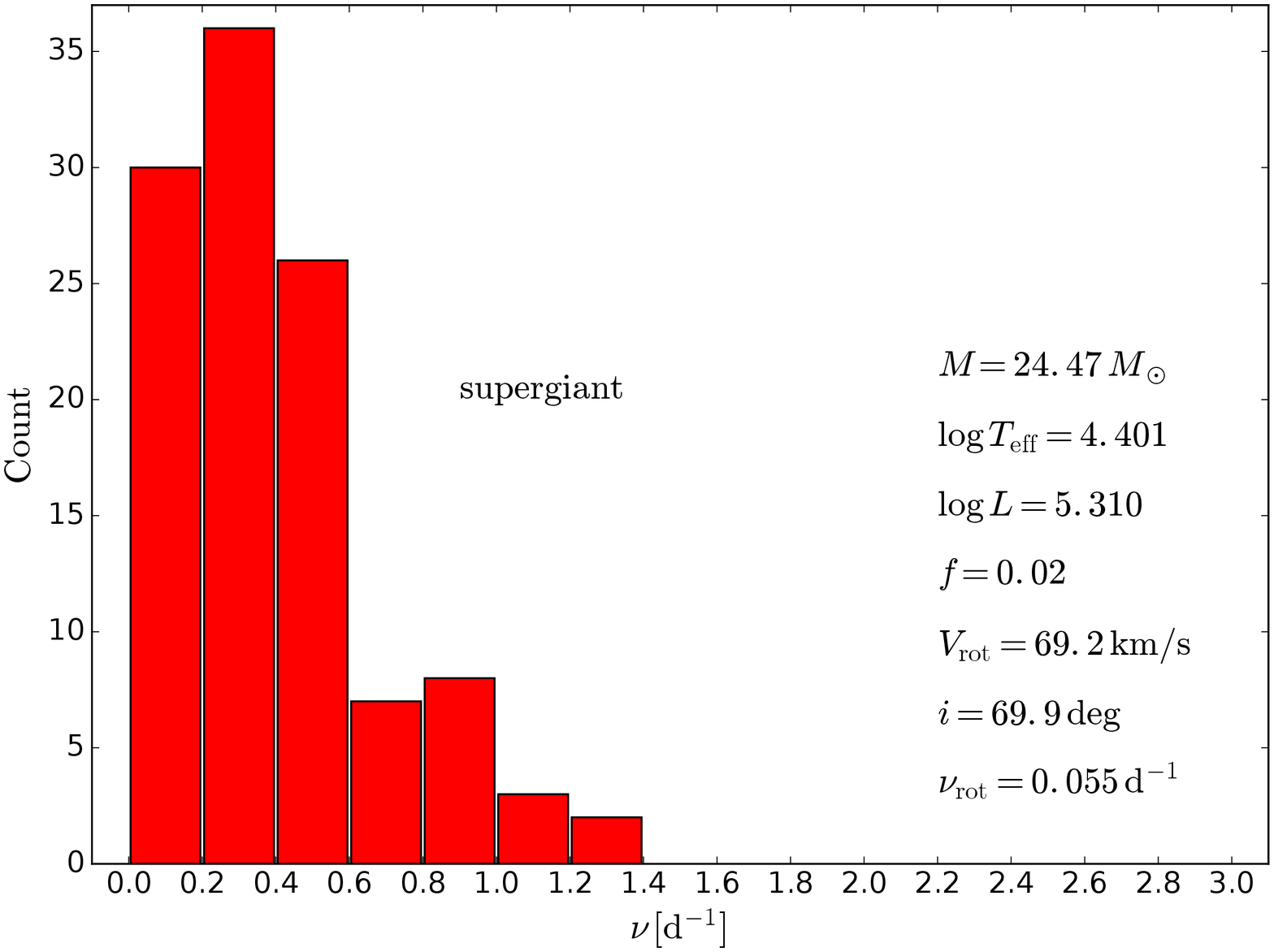}
  \caption{Histogram for the observed frequencies of HD\,163899 (the top-left panel) and the theoretical histograms calculated for main sequence models with the initial masses of $24 M_\odot$ and $22 M_\odot$ (the top-right and the bottom-left panels) and for a supergiant model (the bottom-right panel) with initial mass of $26 M_\odot$. Unstable pulsational modes with the degrees $\ell=0-4$ and the $A_V>0.39$ mmag were considered. The modes were rotational split according to the value of $\nu_rot$ given in each panel.}
\label{fig10}
\end{center}
\end{figure*}

In Fig.\,\ref{fig9}, we show the theoretical values of the photometric amplitudes as a function of mode frequency
for the main sequence model (left panel) and for the supergiant model (right panel). These are the same models used in Fig.\,\ref{fig8}.
As one can see, the criterion $A_V>0.39$ mmag cuts many frequencies but still a large number of them remains.
This can be better estimated if the results are depicted as histograms for the frequencies with the photometric amplitudes $A_V>0.39$ mmag.
In Fig.\,\ref{fig10}, we show the histograms for the observed frequencies (top-left
panel) and for the theoretical ones of the two main-sequence models differing in mass (top-right and bottom-left panels) and a supergiant model (bottom-right panel).
The models in the right panels are the same as considered in Fig.\,\ref{fig8}.

As expected, the rotationally split multiplets populate the gap between the two maxima of $\eta(\nu)$ in Fig.\,\ref{fig8} of the MS pulsational models
and fills the lowest frequency range in both models. 
First of all, we have a much higher number of the theoretical modes than the observed.
This is not a surprise as the result recalls the old problem in the linear theory of stellar pulsations, namely, 
we observe significantly fewer modes than we can predict. An example from the MOST observations is HD\,163830, the SPB star
with 20 frequencies detected comparing to hundreds of unstable theoretical modes \citep{2006ApJ...642L.165A}.

Moreover, a sudden increase of frequencies towards the lower frequencies occurs for both MS and supergiant models.
Thus, even reducing the number of frequencies, we cannot reproduce the shape of their distribution.
The highest observed frequencies also pose a challenge; there are 7 peaks above $\nu = 1.8$ d$^{-1}$.

There can be a few reasons for that. Firstly, we do not know the mode selection mechanism and our simple random selection 
of the values of $\varepsilon$ can be very likely incorrect. Besides, our
crude approximation about the value of the intrinsic mode amplitude, $\varepsilon$,
can be incorrect; not necessarily a frequency peak with the highest photometric amplitude
must have the highest intrinsic amplitude.
Secondly, maybe there are still some inaccuracies in computations of stellar structures resulting from
uncertainties in calibrations of the free parameters describing e.g., semiconvection or efficiency of rotational mixing.
The other source of uncertainties arise from the opacity data.
It is well known that models with the standard opacities cannot account for pulsations in some early B-type main sequence stars
(Daszynska-Daszkiewicz et al. 2016, submitted).
It is also possible that some other mechanism, like stochastic process, is responsible for the light variations of HD\,163899, e.g., subsurface convection.
Finally, the frequencies may originate from a lower mass companion if the star is a binary.

\section{Conclusions} \label{conclusions}

The goal of this paper was to revise the evolutionary status of HD\,163899 using the new determinations of its basic stellar parameters. For the first time, it was possible to determine $\log T_\mathrm{eff}$, $\log L/M$ and rotational velocity of the star with high precision. The new values of these parameters suggest that the star is more luminous, hotter and more massive than it was previously thought. In all former papers treating HD\,163899 it was assumed that the star was beyond the main sequence, either during shell hydrogen burning or during core helium burning phase on the blue loop. That was a valid premise because of the available determination of the spectral type and also because the expected mass of the stars was lower. The most obvious result related directly to the current higher mass estimations is the presence of main-sequence models in the error box of HD\,163899.
The main-sequence models in the error box are rather evolved and close to TAMS. 
They also require a mixing mechanism that leads to broadening of the main sequence.
We found that a higher value of convective overshooting parameter ($f > 0.02$) is needed to obtain MS models in the error box. However, rotation is a key ingredient in our models. Not only is it necessary to reproduce the observed rate but the rotational mixing also greatly helps to deal with various problems related to modelling of massive stars.

The analysis of the HARPS spectra led to the conclusion that HD\,163899 is a nitrogen rich star. The measured nitrogen excess is a clear manifestation of efficient mixing in this star, which we have successfully reproduced by invoking rotational mixing. The observed value of the equivalent ratios, W(N)/W(C) and W(N)/W(O), allowed to estimate the rotation to be about 25\% to 30\% of the critical value of the angular velocity.

We were not able to reproduce the distribution of the observed frequencies with standard pulsational models. 
Possibly our simple approach to modelling the frequency distribution is not adequate
and does not compensate the lack of a mode selection mechanism.
Moreover, there are still many uncertainties in evolutionary models concerning mixing processes as well as opacity data.
Consequently, pulsational models can be incorrect.
Other scenarios involve stochastic processes which can cause the light variations of HD\,163899.
Finally, some features in the spectra may be evidence of a binarity. Given a high percentage of binaries amongst B and O-type stars (e.g., \citealt{2012Sci...337..444S,2014ApJ...782....7D}),
it is very likely that this is the case. This fact could explain also the observed frequencies if they would come from a lower mass companion.

To decide which explanation is true or more likely, further observations and analysis of HD\,163899 are needed.
In particular, a time-series spectroscopy could bring a breakthrough in the interpretation of the variability of the star. Given the fact that main sequence models can also represent HD 163899, this star may not be the prototype of the SPBsg class, which encourages additional search in the ongoing or upcoming space missions such as K2, TESS and PLATO 2.0.
 

\acknowledgments
Based on the archive "HARPS spectra of CoRoT targets", prepared in the framework of the FP7 project n 312844 \textit{SpaceInn - Exploitation of Space Data for Innovative Helio- and Asteroseismology}. This work was financially supported by the Polish NCN grants 2015/17/B/ST9/02082 and 2013/09/N/ST9/00611. Calculations have been carried out using resources provided by Wroclaw Centre for Networking and Supercomputing (http://wcss.pl), grant No. 265.


\end{document}